\documentclass[sigconf]{acmart}

\AtBeginDocument{%
  \providecommand\BibTeX{{%
    \normalfont B\kern-0.5em{\scshape i\kern-0.25em b}\kern-0.8em\TeX}}}

\copyrightyear{2023} 
\acmYear{2023} 
\setcopyright{acmlicensed}\acmConference[CHCHI 2023]{Chinese CHI 2023}{November 13--16, 2023}{Denpasar, Bali, Indonesia}
\acmBooktitle{Chinese CHI 2023 (CHCHI 2023), November 13--16, 2023, Denpasar, Bali, Indonesia}
\acmPrice{15.00}
\acmDOI{10.1145/3629606.3629622}
\acmISBN{979-8-4007-1645-4/23/11}

\usepackage{float}
\usepackage{graphicx}
\usepackage[english]{babel}
\usepackage{blindtext}
\usepackage{enumitem}
\usepackage{multirow}
\usepackage{xcolor}
\usepackage{makecell}
\usepackage{textcomp}

\usepackage{tabularx}

\begin{document}

\title[OperARtistry]{OperARtistry: An AR-based Interactive Application to Assist the Learning of Chinese Traditional Opera (Xiqu) Makeup}

\author{Zeyu Xiong}
\authornote{Both authors contributed equally to this research.}
\orcid{0000-0002-3652-1890}
\affiliation{%
\institution{Computational Media and Arts Thrust}
  \institution{The Hong Kong University of Science and Technology (Guangzhou)}
  \city{Guangzhou}
  \country{China}
}
\email{zxiong666@connect.hkust-gz.edu.cn}

\author{Shihan Fu}
\authornotemark[1]
\orcid{0009-0005-3019-6600}
\affiliation{%
  \institution{Computational Media and Arts Thrust}
  \institution{The Hong Kong University of Science and Technology (Guangzhou)}
  \city{Guangzhou}
  \country{China}
}
\email{sfu663@connect.hkust-gz.edu.cn}

\author{Mingming Fan}
\orcid{0000-0002-0356-4712}
\affiliation{%
  \institution{Computational Media and Arts Thrust}
  \institution{The Hong Kong University of Science and Technology (Guangzhou)}
  \city{Guangzhou}
  \country{China}
}
\affiliation{%
  \institution{The Hong Kong University of Science and Technology}
  \country{Hong Kong SAR, China}
}
\authornote{Corresponding Author}
\email{mingmingfan@ust.hk}

\renewcommand{\shortauthors}{Xiong, Fu and Fan}

\begin{abstract}
  Chinese Traditional Opera (Xiqu) is an important type of intangible cultural heritage and one key characteristic of Xiqu is its visual effects on face achieved via makeup. However, Xiqu makeup process, especially the eye-area makeup process, is complex and time-consuming, which poses a learning challenge for potential younger inheritors. We introduce OperARtistry, an interactive application based on Augmented Reality (AR) that offers in-situ Xiqu makeup guidance for beginners. Our application provides a step-by-step guide for Xiqu eye-area makeup, incorporating AR effects at each stage. Furthermore, we conducted an initial user study (n=6) to compare our approach with existing video-based tutorials to assess the effectiveness and usefulness of our approach. Our findings show that OperARtisty helped participants achieve high-quality eye-area makeup effects with less learning time.
\end{abstract}





\maketitle

\section{Introduction}
Chinese traditional opera, also known as Xiqu, is listed on the Intangible Cultural Heritage (Beijing Opera, Cantonese Opera, Kunqu, Tibetan Opera, and
Shadow Puppets) of UNESCO~\cite{Browseth50:online}. Prior work has investigated ways to preserve~\cite{10.1145/3306214.3338566} and promote~\cite{10.1145/2583114, 10.1145/3290605.3300459} Xiqu. It often takes years of training to hone their skills and effectively portray the character traits of the roles they play\cite{Kuian2006}. According to~\citet{pang2005re} and \citet{wichmann1990tradition}, characters are easier for the audience to recognize when their appearance and clothing are exaggerated. This makes Xiqu art, even if it is a form of music, quite demanding in terms of visual effects.~\citet{li2021contact} noted that in order to ensure the quality of the visual effects, Xiqu Cosmetics frequently employs \textit{grease paint}, which is quite different from everyday cosmetics and frequently contains ingredients that are detrimental to the face~\cite{wang2023insights, wang2020heavy}. At the same time, the process of applying face makeup is time-consuming and complex, as explained by~\citet{liu1997art}. The performers use colorful paintings to paint a variety of symbols and line art on their faces, which emphasize the characteristics, positions, ages, and provenances of the characters. Xiqu face-painters must take into account the individual shape and features of their face upon which they are working \cite{wang2013research}. Moreover, the choice of facial colors and the direction of the lines is complex. Furthermore, Xiqu makeup kit, which includes brushes, water-based paints, powder, and oil-based paints \cite{wang1984Face}, differs from traditional makeup kits. Thus, for beginners who are unskilled in the Xiqu make up techniques, they often need to invest significant efforts to learn them. One popular way to learn make up is by watching online make up instructional video tutorials \cite{riboni2017youtube}. Indeed, using online video tutorials is also a main way to learn Xiqu makeup on social media. However, these online tutorials are often non-interactive and cannot take into account the unique facial features of each individual learner. Thus, beginners have to make guesses and trial-and-error when following video tutorials, which are ineffective and can lead to many errors. 

Augmented Reality (AR) allows digital content to be overlaid in the real world, creating an interactive and immersive experience. We chose to incorporate AR into our work to provide Xiqu makeup learners with a more engaging and intuitive way to learn complicated makeup techniques. With AR, learners can see the virtual effect of makeup applied to their own faces in real-time, allowing them to observe potential issues promptly and make adjustments. Moreover, we leveraged AR to provide step-by-step guidance and feedback, making the learning process more interactive and personalized. In sum, by leveraging AR technology in our design, we sought to make Xiqu makeup learning process more accessible and enjoyable.
Specifically, we sought to answer the following research questions (RQs):

\textbf{RQ1}: What challenges are associated with learning Xiqu makeup?\par
\textbf{RQ2}: How to design an AR-based interactive learning approach to help beginners to learn Xiqu makeup that can address the challenges that they encounter?\par
\textbf{RQ3}: How do AR-based interactive learning approach help beginners learn to make Xiqu makeup compared to video tutorials?\par

We first conducted semi-structured interviews with 12 Xiqu artists with varieties of practical experience in Xiqu performance and found the challenges of learning Xiqu makeup, and needs for detailed and personalized makeup instructions. Based on the findings, we derived four design considerations and designed an interactive AR-based app, OperARtistry, which provides Xiqu makeup beginners with real-time polylines and step-by-step instruction features to better learn the makeup process. We picked the eye area, which is one of the important regions of the face, for this initial evaluation. We recruited 6 participants, who were interested in Xiqu makeup and had different levels of daily makeup skills, to participate in a comparative study, where they used both our AR-based approach (i.e., OperARtistry) and a traditional video tutorial approach. Results show that OperARtistry helped participants achieve higher quality makeup results in terms of color, shape, completion time, effect, and similarity, compared with baseline. Overall, this work lays the foundation for further exploration and advancements in Xiqu makeup guidance. Moreover, the insights gained from our work have the potential for broader generalizability beyond the context of traditional Xiqu makeup. They can be extended to various artistic fields that encompass complex visual effects, opening up new possibilities for enhancing makeup techniques and aesthetics.

\section{Background and Related Work}
Our work has been informed by the prior work investigating \textit{Xiqu Face Painting Background} and \textit{Assistive Technologies for Doing Makeup}.
\subsection{Xiqu Face Painting Background}
In Xiqu facial decoration, there are two main techniques used: make up technique and face painting technique \cite{delza1971picture}. The makeup technique is used to enhance the natural appearance of the face by intensifying desired pink and white skin tones or accentuating features with exaggerated colors. However, it is important that the face still appears natural, and only minimal touch-ups are made to the eyes and eyebrows to make them appear more realistic. On the other hand, the face painting technique is used to create a painted picture of the face, which can sometimes result in the effacement of actual facial features. For example, eyes may be drawn on the forehead or cheeks, and noses and mouths may be distorted. The use of these two techniques depends on the character traits that the Xiqu actors need to convey.

The performers use colorful paintings to paint a variety of symbols and line art on their faces, which emphasize the characteristics, positions, ages, and provenances of the characters. In Xiqu, the symbolism of the painted face is typically classified based on color and configuration \cite{liang1980artistic}, with six primary and six secondary colors, in addition to gold and silver, generally acknowledged in facial painting. The painted face configuration is complex, comprising numerous, at least over 500, painted face characters, excluding regional variations, with major facial delineations consisting of eyebrows, eye sockets, nose, mouth, and forehead. The combination of color and shape in each of these areas not only suggests the image of a historical personage but also captures the personality of a Xiqu character.

\begin{figure}[tbh!]
    \centering
    \includegraphics[width=0.9\linewidth]{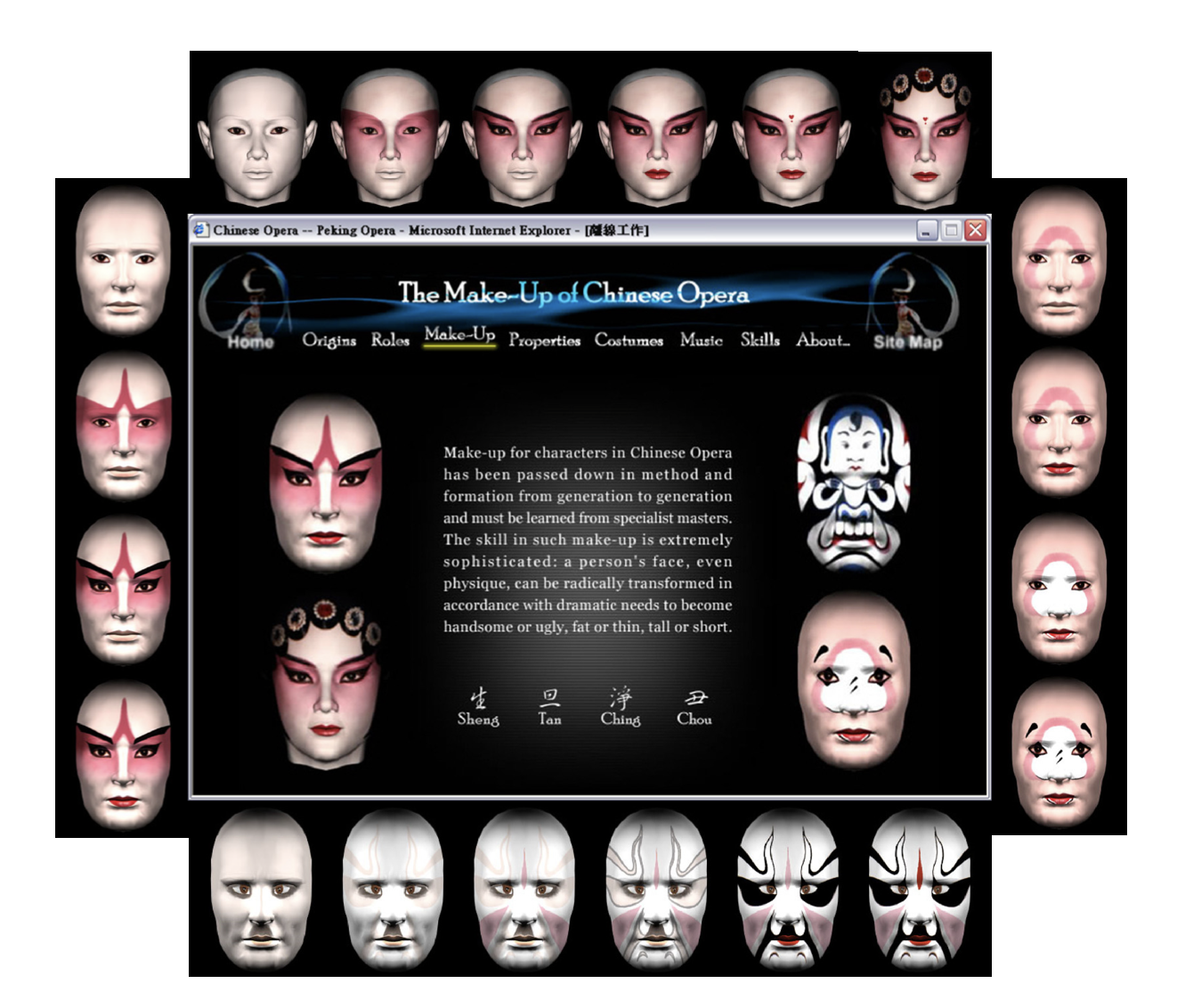}
    \caption{The make up Steps of the Four Main Roles in Chinese Opera~\cite{wang2013research}, Left: Sheng, Top: Dan, Right: Chou, Buttom: Ching}
    \label{fig:make up steps}
\end{figure}

The Xiqu character's make up process involves multiple steps that use different dyes and brushes to apply the colors, as shown in the make up Steps of the Four Main Roles in Chinese Opera (see Figure~\ref{fig:make up steps}). For the Jing role, Kao Teng, the makeup processes are:
    1) A base of white powder is applied, and black powder is applied around the eyes to prevent oil-based makeup from seeping into them during a performance; additional black powder is applied under the nose to prevent the beard from slipping off the actor’s face.
    2) A brush dipped in white paint is used to draw the approximate positions of the eyebrows, eyes, nose, and facial patterns.
Item: Black oil-based paint is applied to unpainted areas.
    3) Another layer of white paint is added to set the facial pattern in place.
    4) The pink powder is dusted onto the cheeks over the white paint.
    5) A single line of oil-based red paint is applied to the forehead~\footnote{http://jinju.koo.org.tw}.


Among the major facial delineations in Xiqu makeup, the \textbf{\textit{eye-area}}, including eye sockets and eyebrows, is particularly important in conveying the emotions and personalities of the characters. The shape and color of the eyebrows, for instance, can suggest the age, gender, and temperament of the character, while the eye sockets can convey the character's mood and intentions. As such, mastering eye-area makeup techniques is crucial for learners to accurately portray Xiqu characters on stage.

Furthermore, the predominant approach to learning Xiqu makeup through social media relies on video instructions. However, online makeup tutorials often lack interactive features that consider individual differences in skin and face format, resulting in variations in product application. This limitation becomes evident as video tutorial presenters may not be well-versed in addressing the unique needs of each user.

Due to the complexity of Xiqu makeup and the lack of interactive guidance in online tutorials, many learners struggle to achieve the desired results, especially in the eye area. Therefore, we are motivated to explore an interactive approach to making Xiqu makeup learning process more accessible for beginners, for example, by providing learners with personalized guidance and real-time feedback.


\subsection{Assistive Technologies for Doing Makeup}


\subsubsection{Makeup try-on \& Alternatives}

Many studies have been conducted in order to propose technical improvements for assisting makeup or simulating makeup facial try-on effects. \citet{iwabuchi2009smart} presented Smart Makeup Mirror, a system that assists the makeup process using a high-resolution camera above a screen display, with functions including displacing face from different angles and providing different lighting conditions. \citet{treepong2018makeup} presented an interactive face makeup system to provide a virtual
makeup experience, consisting of a face painting application that projects makeup effects onto the face and a set of touch-detectable makeup tools that enable users to feel the experience of real makeup. The iMake is a computer-aided eye makeup design system developed by \citet{nishimura2014imake}, which creates eye makeup designs using images chosen by the user, moreover, these designs can be previewed in an augmented simulation system and can be transferred onto the user's eyelids using printed transfer sheets. \citet{kao2016chromoskin} proposed a dynamic color-changing make up system that allows the wearer to seamlessly change their appearance. Specifically, they have designed an interactive eye shadow tattoo prototype consisting of thermochromic pigments that allow the user to achieve a color-changing banding effect. However, due to the uniqueness of the Xiqu makeup, there is no evidence yet that these alternatives have been applied to the traditional Xiqu makeup, and traditional oil paints are still widely used to achieve desired visual effects in Xiqu.

\subsubsection{Makeup aid for specific scenes \& user group}

In the HCI community, researchers have tried various ways of making the makeup aid more accessible and interactive. For example, \citet{10.1145/3411764.3445721} has created a system for automatically generating hierarchical makeup video tutorials. The study of facial makeup for specific scenes, like theater drama, has also been conducted. \citet{cai2010real} presented a makeup design support system for the Peking Opera, which combines the eye, nose, and mouth designs to generate new designs. Users can view these designs on a 3D-rendered model. Moreover, prior work has also investigated technologies for assisting people with disabilities with make up. For example, \citet{li2022feels} suggested the practices and challenges of makeup for people with visual impairments, and interactive feedback is mentioned as one of the key points to making the makeup process more accessible. Additionally, researchers have investigated makeup-wearing prediction. For example, \citet{10.1145/3491102.3517659} has developed an algorithm that provides visual aids for makeup product selection. To our knowledge, how to design an interactive approach to help people better learn Xiqu makeup with less effort remains to be an open research question.

\subsubsection{Augmented reality system}

Regarding Augmented Reality (AR) techniques, there are lots of work and applications for facial try-on, for example: (1) AR makeup try-on~\cite{borges2019virtual, marelli2022designing, 10.1145/3283289.3283313}, (2) AR makeup creativity enhancement \cite{treepong2018makeup}, (3) AR face masks \cite{premarathne2022dc, liu2020comparing, 10.1145/3599609.3599635, 10.1145/3025453.3025722}, etc. Such techniques typically require fundamental facial detection algorithms to extract facial feature points \cite{kumar2019face}. Moreover, they are often designed to demonstrate makeup effects rather than in-depth makeup tutorials. Furthermore, fewer, if any, AR applications are designed for opera face painting. The existing virtual makeup tools have also been
 embedded in the special effects of certain retouching apps (e.g. MeiTuXiuXiu~\footnote{https://pc.meitu.com/}, Douyin~\footnote{https://www.douyin.com/}, etc), however, these applications are mainly for entertainment and do not reach the requirements for Xiqu lovers as well as makeup learners. Therefore, even though AR make up is widely used, there are few systems made to incorporate the characteristics of Xiqu make up, and for Xiqu enthusiasts to participate in AR-enabled theatre make up. Thus, in this work, we sought to address this gap by designing and evaluating AR-based interactive app to assist beginners to learn Xiqu makeup.

\section{FORMATIVE STUDY}
To address RQ1, we conducted a formative study that involved semi-structured interviews with 12 Xiqu artists. The primary objective of this study was to gain initial insights into the experiences of learning Xiqu makeup, with a specific focus on the challenges encountered during the makeup process. The findings from this formative study highlight the user needs and preferences, which served as valuable guidance for the design of a prototype aimed at assisting beginners in learning the art of Xiqu makeup.

\begin{table*}[ht]
\caption{Basic demographic information of the participants in semi-structured interviews}
\begin{tabular}{cccccc}
\hline
\multicolumn{1}{l}{Id} & \multicolumn{1}{l}{Gender} & \multicolumn{1}{l}{Age} & \multicolumn{1}{l}{Years of Xiqu Makeup Experience} & \multicolumn{1}{l}{Opera Type} & \multicolumn{1}{l}{Avg  Face Paint Wearing Frequency} \\ \hline
1                      & M                          & 15-18                                  & 6-10                                                & Cantonese opera                & 2-3 times / week                                     \\
2                      & F                          & 26-35                                       & 1-5                                                 & Beijing Opera                  & 1-2 times / month                                    \\
3                      & F                          & 15-18                               & \textgreater{}10                                    & Beijing Opera                  & 1-2 times / month                                    \\
4                      & M                          & 18-25                                         & 1-5                                                 & Beijing Opera                  & 1-2 times / month                                    \\
5                      & F                          & 18-25                                         & 6-10                                                & Yue Opera                      & \textgreater 3 times / week                          \\
6                      & F                          & 18-25                                          & 1-5                                                 & Kunqu                          & once a week                                          \\
7                      & F                          & 18-25                                            & 1-5                                                 & Wuxi Opera                     & once a week                                          \\
8                      & F                          & \textgreater{}55                              & \textgreater{}10                                    & Beijing Opera                  & 2-3 times / week                                     \\
9                      & F                          & 36-45                                       & \textgreater{}10                                    & Cantonese opera                & once a week                                          \\
10                     & F                          & 15-18                                  & 6-10                                                & Cantonese opera                & 1-2 times / month                                    \\
11                     & M                          & 36-45                                          & \textgreater{}10                                    & Kunqu                          & \textgreater 3 times / week                          \\
12                     & M                          & 26-35                                       & 6-10                                                & Henan Opera                    & 1-2 times / week                                     \\ \hline
\end{tabular}
\label{table: meta1}
\end{table*}

\subsection{Semi-structured Interviews}
We conducted online semi-structured interviews with 12 Xiqu artists (4 male, 8 female), with a wide variety of practical experience in Xiqu performance (See Table~\ref{table: meta1}). Each interview session lasted about one hour and every participant was paid in accordance with local standards. The interview focused on \textit{how they learned to apply Xiqu makeup}, \textit{the steps to apply Xiqu makeup}, \textit{specific tools}, and \textit{cosmetics}. To analyze the interview data, we followed the thematic analysis~\cite{oktay2012grounded} method. The open-coding procedure was carried out independently by the two co-authors once they had familiarized themselves with the data by reading the interview transcripts. Deductive and inductive coding techniques were combined, and we set up themes of "practices, challenges" for the coding process. To create a unified codebook, the two coders regularly discussed the codes and worked out their differences. All co-authors were invited to additional meetings to reach agreements based on the initial coding results. We summarize our tags related to the two themes in the research questions at the end. 

\subsection{Findings}
We present our findings from the formative study in this section and highlight particular features of the Xiqu makeup.
\subsubsection{Challenges faced when learning make up.} Even with in-person teaching from trained instructors, we found that novices can find it challenging to completely master the art of Xiqu makeup, while this kind of instruction has limited access for a wide range of learners. All these findings highlight the need to design a makeup tutorial app for them.
\paragraph{Importance of makeup order.}  All of our participants described their Xiqu makeup routines in a specific order, in contrast to daily makeup routines, which have no predefined order. This is due to the unique nature of Xiqu cosmetics, which require that the oil painting be applied before the eyeshadow powder. If not, the makeup will come off easily and the color effect will be unsatisfactory (\textit{P3: " Applying the oil painting first helps create a smooth base for the eyeshadow to adhere to, also, following a step-by-step approach ensures that each component of the Xiqu makeup is applied correctly and maximizes the lasting effects of the makeup"}). This leads to our first design consideration~\textbf{DC1: The Makeup tutorial should be presented step by step.} 
\paragraph{Difficulty of eye makeup. }The majority of our participants (N=10) rated applying makeup to the eye area as the most challenging step in the entire makeup process. This is due to a number of factors, such as the sunken eye structure and the need for extreme symmetry on the eye part (\textit{P8: "Creating symmetrical eyeliner in the precise area requires much practice and patience. Eye makeup, in particular, is the most intricate part of the process due to its multi-layered nature and the need for various shades of cosmetics. On the other hand, the rest of the face simply requires applying oil-based paints without any specific difficulties"}). This derived~\textbf{DC2: The Makeup tutorial should offer more help for eye makeup.}

\subsubsection{Needs for detailed and personalized Makeup instructions} 
During our interviews, we discovered that the artistry of Xiqu make up extends beyond the selection of appropriate cosmetics. It also involves the consideration of each individual's distinct facial features to make necessary adjustments. This finding offers valuable insights and design considerations for developing our system with detailed and personalized makeup instructions.

\paragraph{The variety and complexity of cosmetic tools.} Xiqu cosmetics, which use distinctive brushes and colors, are different from common cosmetics. While the proper use of brushes can make the makeup process more effective, the color choice of cosmetics is the key to present Xiqu characters' personalities (\textit{P11: "I would adjust the color on my face to fit the character}). As a result, we should standardize the cosmetics accordingly - 
\textbf{DC3: The Makeup tutorial should specify cosmetic tools and color choices.}

\paragraph{The adjustment of makeup for different face features.} Some of our participants (P3, P4, P8) mentioned that they customize their Xiqu makeup based on their unique facial features, Specifically, they emphasized that Xiqu makeup is not fixed and, moreover, may require adjustments to enhance symmetrical or larger eyes (\textit{P3: "I need to magnify my eyes with makeup to make them more attractive"}). Therefore, it is essential to adjust makeup tutorials according to individual features - 
\textbf{DC4: The application should pay attention to each person's individual facial features.}

\section{System Design}
\begin{figure*}[tbh!]
    \centering
    \includegraphics[width=0.8\linewidth]{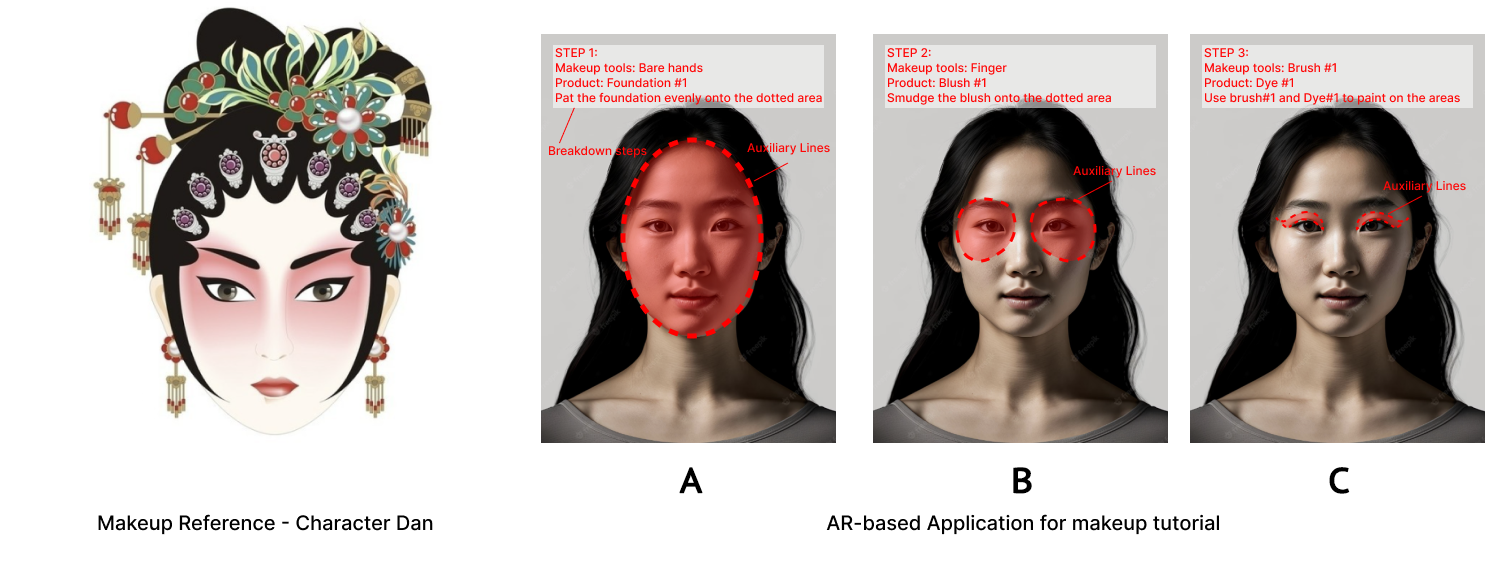}
    \caption{System Design: AR Opera Paintings Tutorial. The Left shows a makeup reference (intended completion), the Right shows interactive tutorial sample steps, and each step includes text-based instructions (including makeup tool \& powder selection advice) and real-time auxiliary lines. Specifically: (A) Step1, pet the foundation evenly onto the whole face with bare hands, (B) Step2, smudge the blush onto the eye area with fingers, (C) Step3, use brush 1 and dye 1 to paint the eyeliner.}
    \label{fig:design-2}
\end{figure*}

Informed by the formative study findings and the derived DCs, we designed an interactive AR-based app to assist Xiqu enthusiasts with their facial make up efforts. Our app focuses on assisting with eye area make up~\textbf{(DC2)}. The app provides instructions in a step-by-step format~\textbf{(DC1)}, accompanied by corresponding auxiliary lines and coloring instructions, which are applied progressively on the face. Such an approach affords a more intuitive and precise make up guide. Also, The matching cosmetics are identified in the application and are clearly numbered~\textbf{(DC3)}. The AR technology offers customized makeup tutorials by locating certain facial points, taking into account each user's individual facial feature~\textbf{(DC4)}. The procedure is specified in Figure~\ref{fig:design-2}.

\begin{figure}[tbh!]
    \centering
    \includegraphics[width=\linewidth]{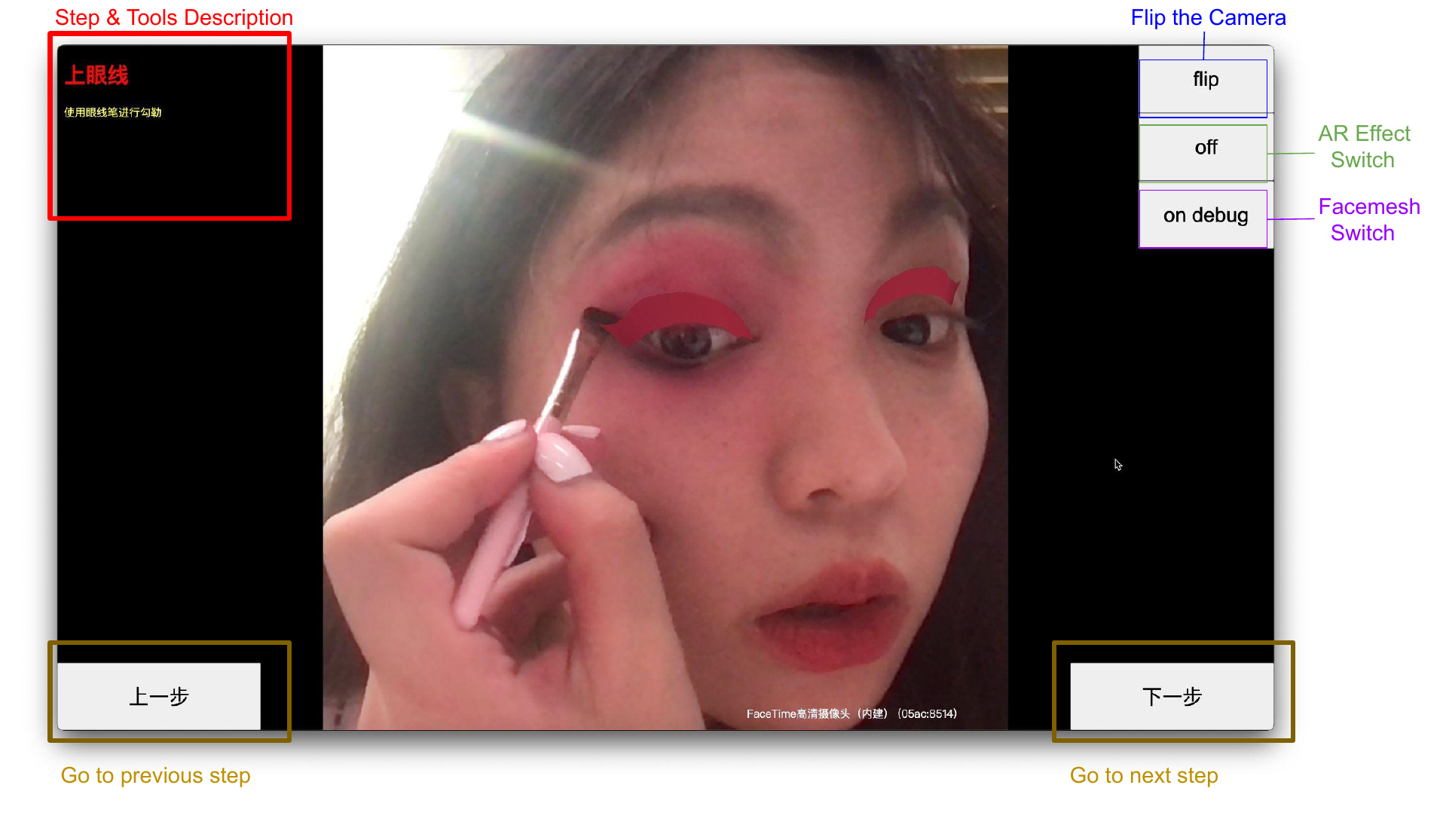}
    \caption{Final Prototype of OperARtistry: Tutorial Page Demonstration. The real-time AR camera is positioned in the central. Top Left shows the step \& tools description. Top Right is the feature hub (flip the camera, ``on / off'' the AR effects, and show debug mode). The bottom has 2 buttons to transfer next or previous steps.}
    \label{fig:detail}
\end{figure}

Beyond the basic design, we also consider the ``flip'' and ``on / off'' functionalities. The ``flip'' function is used to flip the camera, which helps the user check whether the makeup is left-right symmetrical or not, while the ``on / off'' function is used to let users choose whether they want to turn on AR effects in order to see the difference between the assist and themselves in the real mirror. Figure~\ref{fig:detail}\footnote{The blurring effect of the figure is added manually for authorship anonymization} shows the final application's ``tutorial page''. The main camera module is in the center of the interface. Steps \& tool description are on the upper left, while on the upper right, there are 3 buttons for ``flip'', ``on / off'', and ``on debug'' (used to check the face mesh alignment). Two buttons at the bottom are used to switch to the previous/next step of the tutorial.

\begin{figure}[tbh!]
    \centering
    \includegraphics[width=\linewidth]{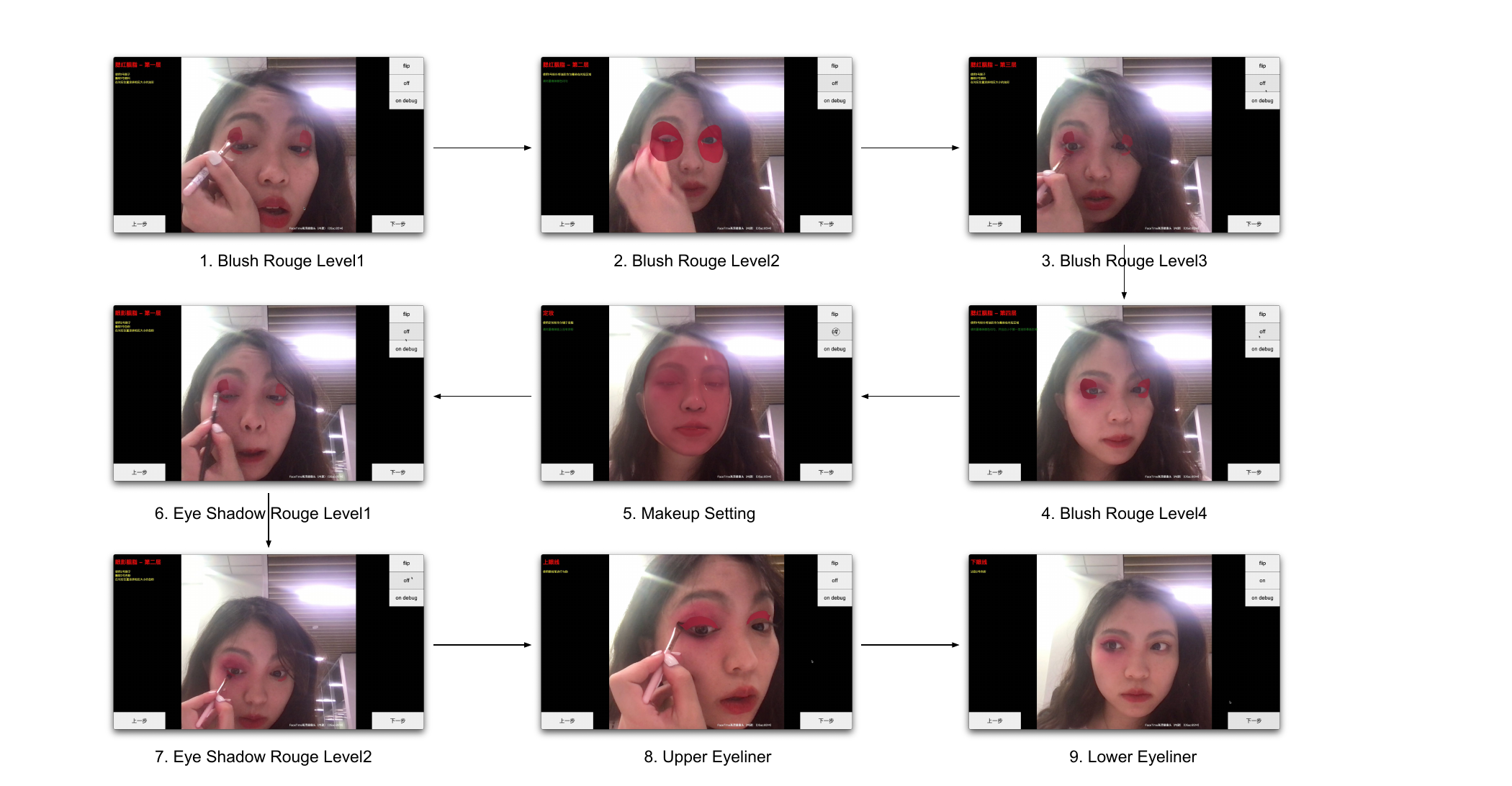}
    \caption{Nine Procedures of Eye Makeup Tutorial. Users need to finish 4 levels of blush rouge, makeup setting, 2 levels of eye shadow rouge, and upper \& lower eyeliner in a row.}
    \label{fig:procedure}
\end{figure}

\begin{figure}[tbh!]
    \centering
    \includegraphics[width=\linewidth]{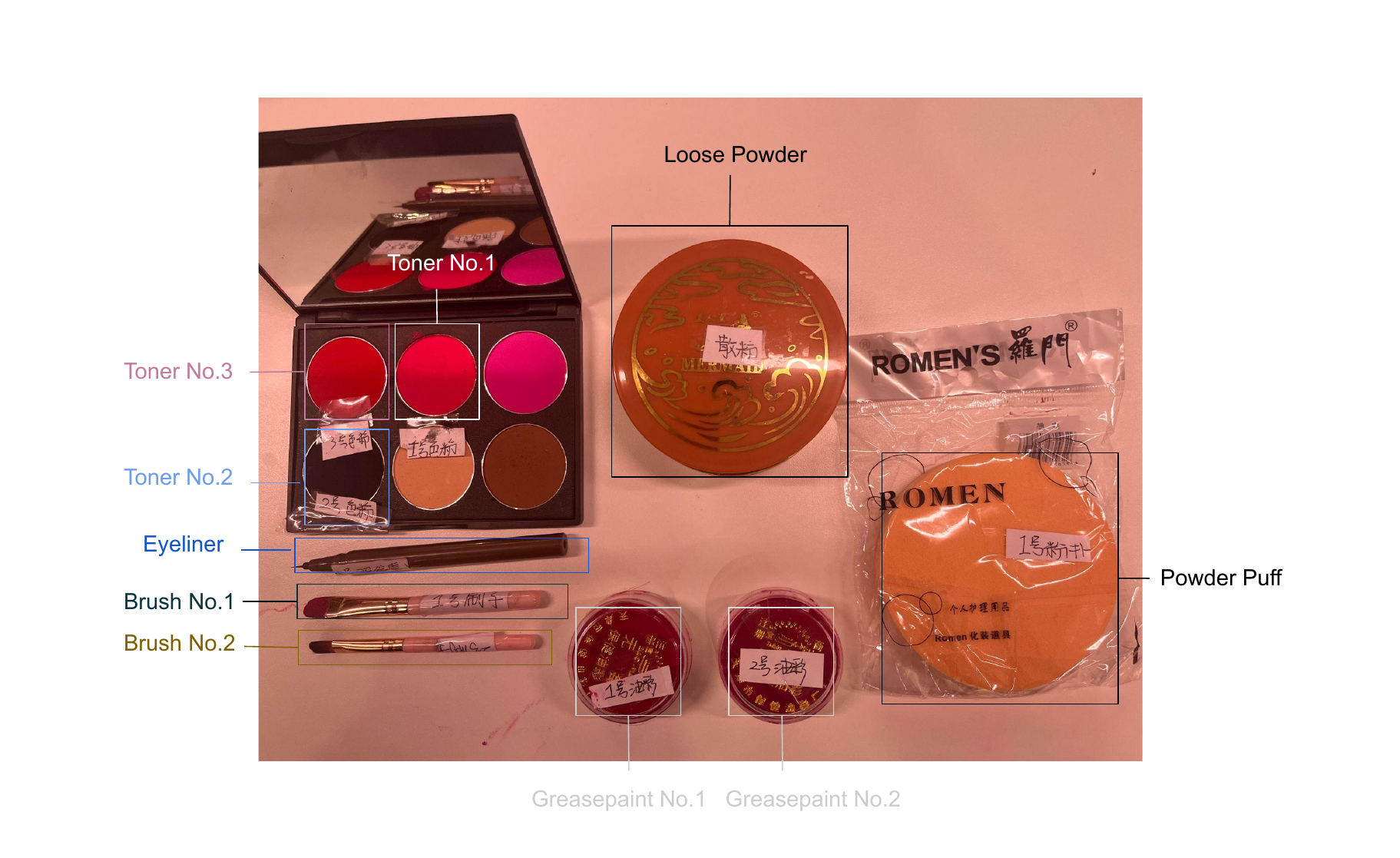}
    \caption{Eye Makeup Tutorial: Tools for Makeup. We provide 3 types of toners, 2 types of brushes, 2 types of grease paints, 1 eyeliner, 1 loose powder, and 1 powder puff for users to complete the tutorial.}
    \label{fig:tools}
\end{figure}

For the content of the tutorial, we mainly focus on ``Eye Makeup'' learning. We have studied the steps of professional makeup artists and broken them down. We finally designed the tutorial for 9 steps (shown in figure~\ref{fig:procedure}) with tools descriptions in detail (shown in Figure~\ref{fig:tools}): 1) \textbf{Blush Rouge Level1} Use \textit{Brush No.1} and \textit{Greaspaint No.1} to fill over the corresponding area. 2) \textbf{Blush Rouge Level2} Use \textit{Powder Puff} to smudge the oil evenly over the corresponding area. 3) \textbf{Blush Rouge Level3}  \textit{Brush No.1} and \textit{Greaspaint No.2} to fill over the corresponding area. 4) \textbf{Blush Rouge Level4} Use \textit{Powder Puff} to smudge the oil evenly over the corresponding area. 5) \textbf{Makeup Setting} Apply \textit{Loose Powder} evenly over the entire face. 6) \textbf{Eye Shadow Rouge Level1} Use \textit{Brush No.2} and \textit{Toner No.1} to fill over the corresponding area. 7) \textbf{Eye Shadow Rouge Level2} Use \textit{Brush No.2} and \textit{Toner No.3} to fill over the corresponding area. 8) \textbf{Upper Eyeliner} Use \textit{Eyeliner} for outlining the upper area of the eye. 9) \textbf{Lower Eyeliner} Use \textit{Eyeliner} for outlining the lower area of the eye.


\section{System Implementation}

\subsection{Overview}

For the system implementation, we build a web application that makes sure all the functionalities work as expected. We refer to some open-sourced materials in the development, for example: (1) DeepAR SDK (\cite{terzopoulos2021comparative})\footnote{https://www.deepar.ai/}, (2) Perfect Crop~\footnote{https://www.perfectcorp.com/business/technologies/makeup-ar}, (3) Aryel~\footnote{https://aryel.io/solutions/virtual-try-on/}, etc. We finally used React.js~\footnote{https://react.dev/} web application framework plus three.js~\footnote{https://threejs.org/} (a 3D library in JavaScript), on top of which we introduced a facial feature point localization algorithm and an enhanced logic step.


\subsection{Facial Feature Points Detection}

To achieve the effect of makeup steps floating on the face, we finish the positioning of the makeup map with the face by recognizing the facial feature points. There are lots of state-of-the-art facial feature point detection methods that have been applied in practice (e.g.~\cite{10096203, xiong2022face2statistics}), and in order for the system to implement the method in real-time and without latency, we made adaptations in google's facemesh model (\cite{ansari20073d, sangal2022drowsy})~\footnote{https://developers.google.com/ml-kit/vision/face-mesh-detection} as well as the tensorflow.js~\cite{smilkov2019tensorflow} framework with the help of webGL rendering technology. Based on the results of facemesh, we can obtain 468 facial feature points simultaneously.
\subsection{Facial Feature Points Alignment}

\begin{figure}[tbh!]
    \centering
    \includegraphics[width=\linewidth]{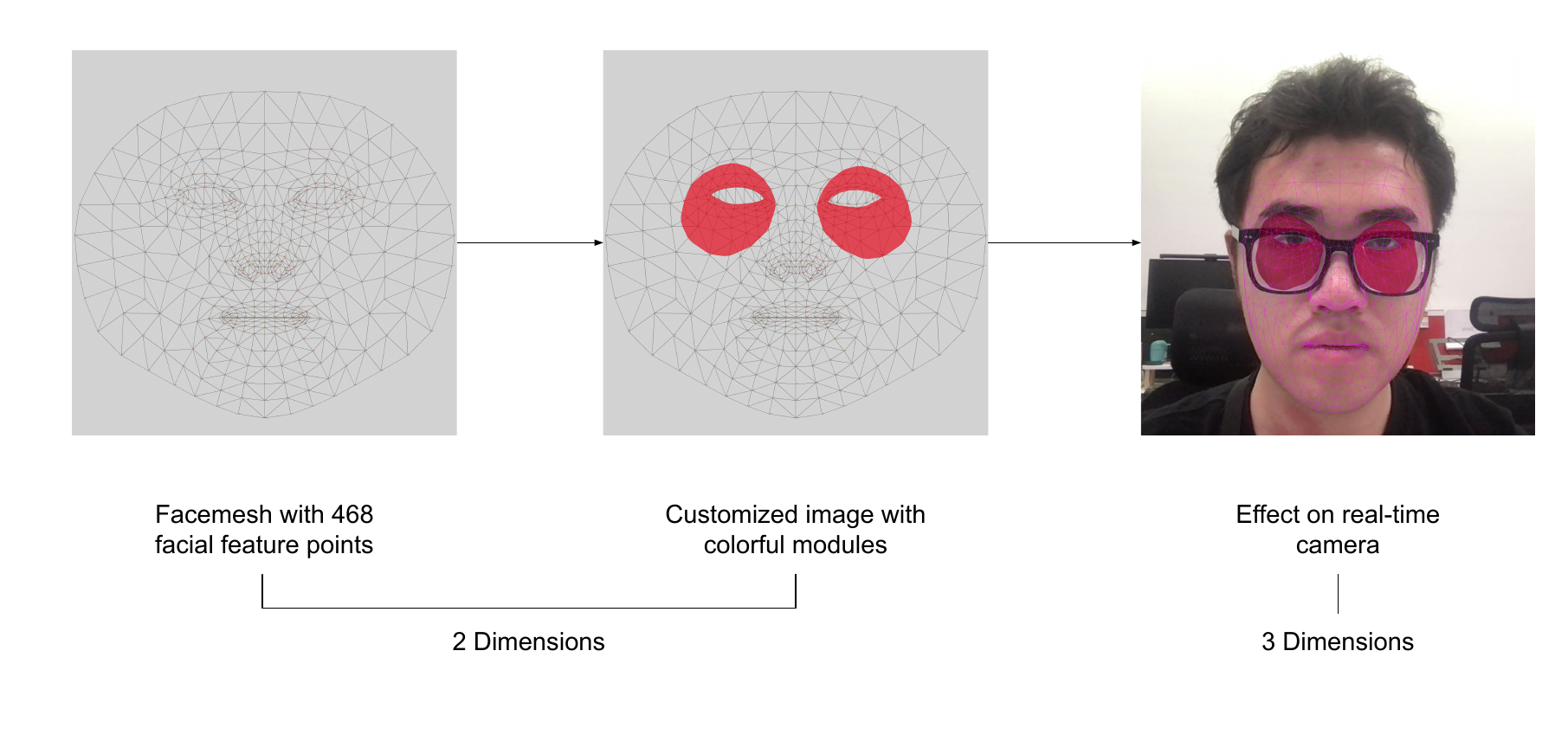}
    \caption{Feature Points Alignment: from 2-Dimension to 3-Dimension. Left: a standard face mesh with 468 facial feature points; Middle: add customized colorful modules in 2-dimension; Right: map 2-dimension face mesh into 3-dimension, and show effects on real-time camera.}
    \label{fig:combined}
\end{figure}

After detecting 468 feature points on the face, We need to match the prepared image to the corresponding feature points. According to Figure~\ref{fig:combined}, we can two-dimensionalize the feature points into an image, based on which we can draw the desired pattern on the image. We then imported the pattern into the system, and the process of changing from 2D to 3D was completed again. In the end, we get the facial effect under the real-time camera.

\section{User Study}

\subsection{Participants}

We recruited 6 participants (2 males, 4 females, aged 22–28) who possess a genuine interest in Xiqu and are enthusiastic about the Xiqu makeup process by posting recruitment messages. To establish a comprehensive understanding of our participants' backgrounds and expertise in makeup, we administered an online recruitment survey prior to the study. This survey allowed us to assess participants' familiarity with make up tools and their prior experience in make up application. Subsequently, we classified participants into three distinct proficiency levels (novices, competent, and proficient) based on their responses in the recruitment survey. This categorization enabled us to examine the influence of varying levels of makeup expertise on the outcomes of our study. For a detailed overview of participants' backgrounds and proficiency levels, please refer to Table~\ref{basicInfo}. 

\begin{table*}[!ht]

    \centering
    \caption{Participants' Background Information and Proficiency Levels}
    \begin{tabular}{lllll}
    \hline
        \textbf{Participant ID} & \textbf{Gender} & \textbf{Age} & \textbf{ Familiarity with make up\textsuperscript{1}}  & \textbf{Proficiency Level} \\ \hline
        P1 & Female & 23 & High & Proficient \\ 
        P2 & Male & 23 & Moderate & Novice \\ 
        P3 & Female & 26 & Low & Competent \\ 
        P4 & Female & 22 & Moderate  & Competent \\ 
        P5 & Male & 22 & Low & Novice \\ 
        P6 & Female & 28 & High & Proficient \\ \hline
    \end{tabular}

    \smallskip
    \begin{minipage}{11.5cm}
        \textsuperscript{1} \begin{small}Proficiency levels were determined based on participants' self-reported familiarity with make up tools and their make up experience, as assessed in the online recruitment survey. \end{small}
    \end{minipage}
    
    \label{basicInfo}
\end{table*}



\subsection{Apparatus and Settings}
We offer participants two kinds of Xiqu makeup tutorials: AR-based make up application (OperARtistry) and video makeup tutorial.
For the AR-based makeup application, we utilized a computer equipped with a built-in camera as the primary device. In contrast, the video makeup tutorial~\footnote{http://xhslink.com/k7cxVp, duration: 19s - 27s} was curated from professional Xiqu makeup artists who have garnered a high recommendation index on popular social media platforms. We meticulously reviewed and selected videos that closely aligned with the makeup steps we had identified during our field research. Specifically, we focused on identifying videos with comprehensive instructions and demonstrations of the eye makeup process, as it is a critical aspect of Xiqu makeup. The selected video tutorials served as standardized references for participants and ensured consistency across the study.

To maintain consistency in the study environment, participants utilized the same computer to access the tutorial content. This approach ensures that participants experience the tutorials in a controlled and uniform manner.

\subsection{Procedure}\label{sec:procedure}

\begin{figure*}[tbh!]
    \centering
    \includegraphics[width=0.8\linewidth]{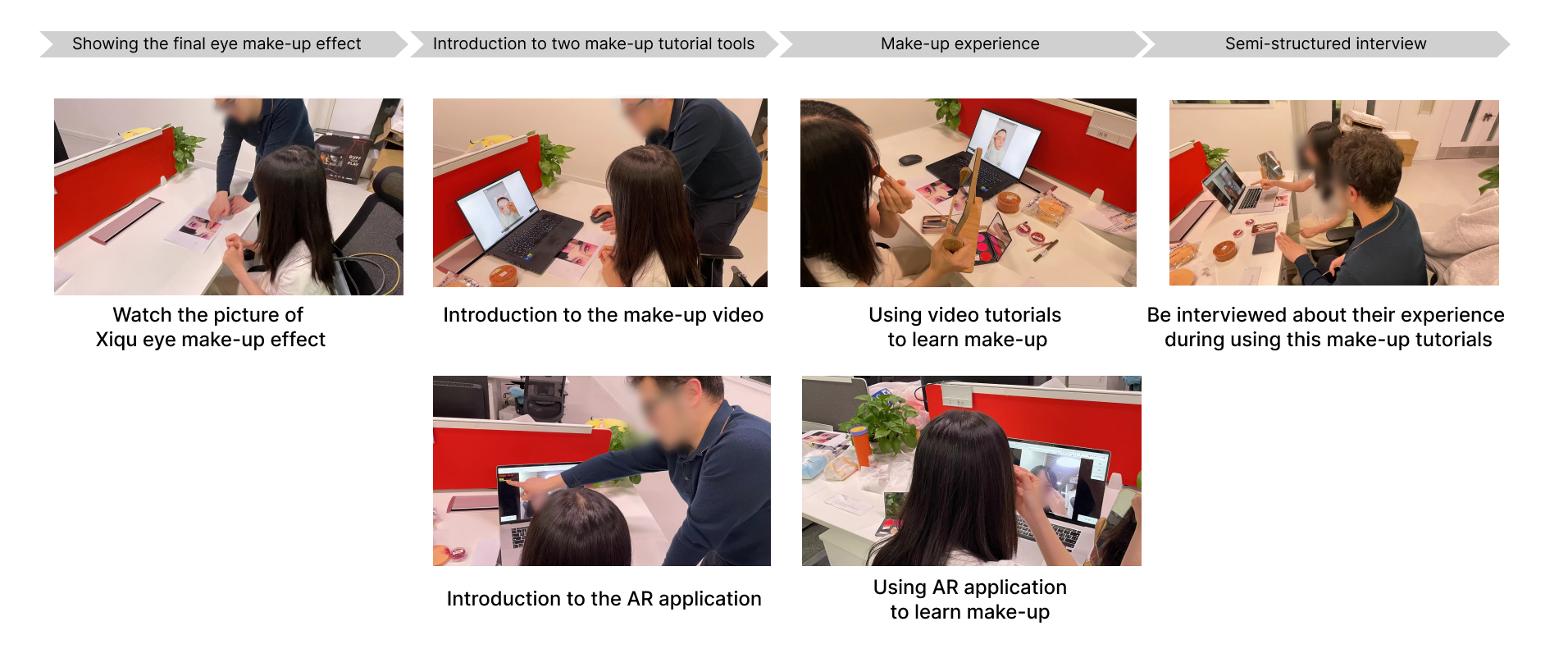}
    \caption{The User study process: (1) showing participants the intended eye makeup effect, (2) introduction session of the makeup video and the AR application, (3) experiencing makeup tutorial by AR application and baseline, (4) a semi-structured interview after experiencing the tutorial.}
    \label{fig:process}
\end{figure*}

The user study consisted of four distinct parts, as illustrated in Figure~\ref{fig:process}. All study procedures were conducted in a dedicated, empty room with the explicit consent of the participants. The experimenter interacted with the participants in a face-to-face setting, ensuring direct engagement and real-time observation of their makeup application process. Each part of the study is described in detail below.

\textbf{Part One.} In this initial phase, participants were presented with a frontal view of a standard Xiqu eye makeup, accompanied by an oblique side view, to provide them with a comprehensive understanding of the desired makeup effect and intricate details. Participants were given the opportunity to continuously refer to this image throughout the makeup application process, allowing them to maintain a clear visual reference.

\textbf{Part Two.} Once participants indicated their familiarity with the Xiqu eye makeup effect, the second part of the study focused on introducing them to the content of the video tutorial and the fundamental functionalities of the AR-based tutorial. Participants were granted autonomy to explore and experience the usage of both makeup tutorial tools, gaining hands-on familiarity with their features and capabilities.

\textbf{Part Three.} After familiarizing themselves with the basic functionalities of the tools, participants proceeded to apply makeup with the assistance of two types of tutorials. The provided makeup tools were used to maintain consistency among participants. In order to minimize any potential order effects, the sequence of the tutorials was presented in different orders to participants using a balanced Latin Square design, which effectively distributed any bias.


\textbf{Part Four.} The objective of this section was to gather detailed insights into participants' experiences and perceptions of the AR-based makeup application in comparison to the video tutorial. First, the user will complete a scale from 1 to 5, from the final makeup's perspective of color, shape, speed, effect, and similarity. Next, a semi-structured interview was conducted with each participant in the same room. Sample interview questions encompassed various aspects, such as overall impressions of the AR makeup experience, utilization of different AR makeup tutorial features, identification of useful and less useful features, and an evaluation of the advantages and disadvantages of AR makeup in relation to video makeup.

\section{Results}
In this section, we present results and evaluations of the system performance and user experience regarding the AR makeup application in comparison to traditional video makeup tutorials. By capturing participants' viewpoints, we gain insights into their overall experiences and evaluations of the AR-based approach.

\subsection{Performance Evaluation}

\begin{figure}[tbh!]
    \centering
    \includegraphics[width=0.6\linewidth]{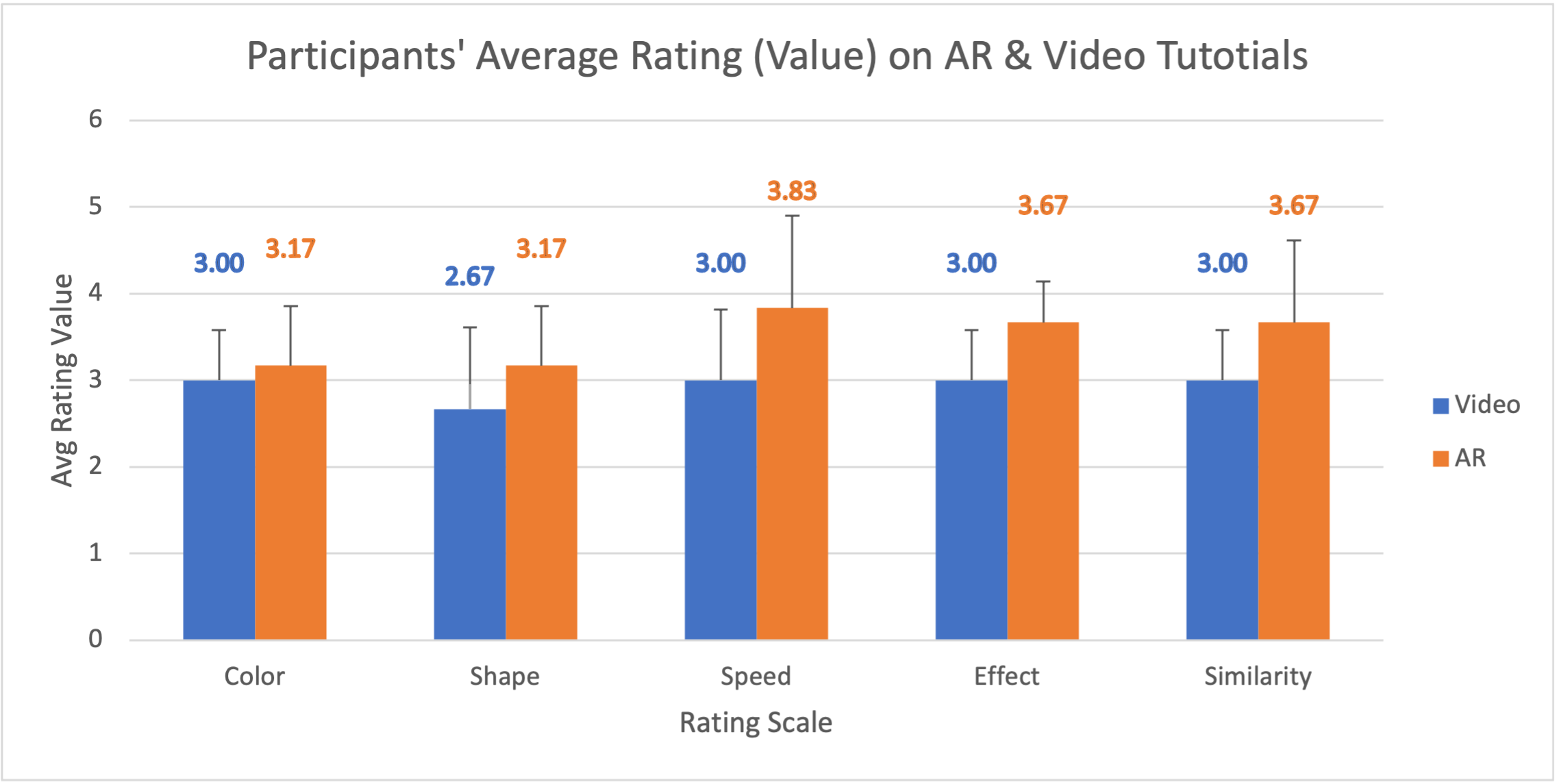}
    \caption{Participants' Avg Ratings on AR \& Video Tutorials (scale: 1 - 5)}
    \label{fig:avg}
\end{figure}

As discussed in section~\ref{sec:procedure}, we collect rating data among several attributes (rating scale 1-5) from the participants. Thus, we summarized these data as a whole, and figure~\ref{fig:avg} shows the detailed average rating data with the error bar (calculate by the standard deviation divided by the square root of the sample capacity) from 6 participants.

\begin{figure}[tbh!]
    \centering
    \includegraphics[width=0.6\linewidth]{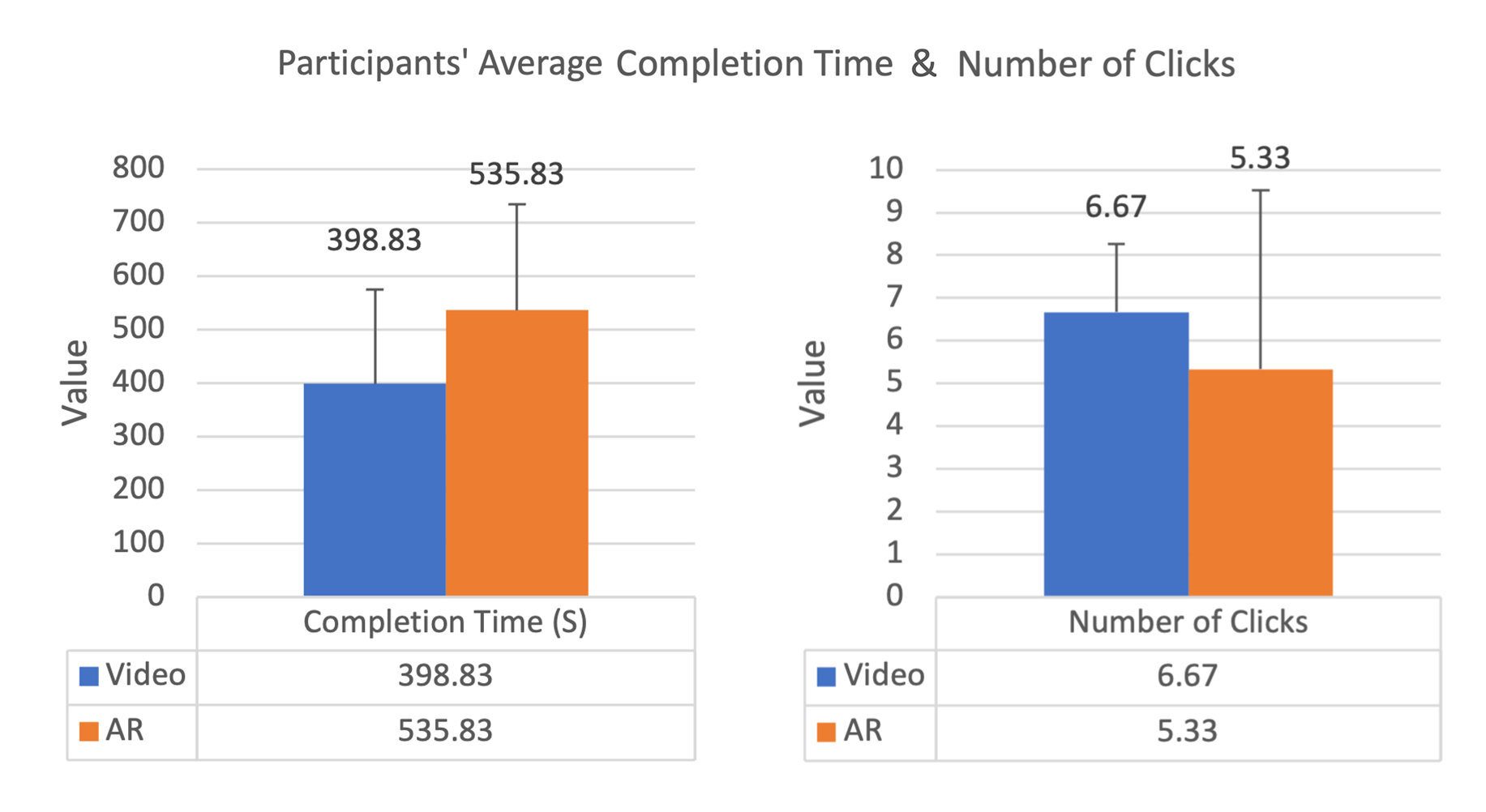}
    \caption{Avg \& Std value of Completion Time \& Number of Clicks}
    \label{fig:acn}
\end{figure}

Besides, we also recorded the number of clicks \& completion time that the participants clicked to the video ``play \& pause'' and AR ``on \& off'' buttons. The details are shown in table~\ref{fig:acn}.




\subsubsection{Finding 1: General Performance -- Victory of AR Tutorial}

Overall, all rating metrics (color, shape, speed, effect, and similarity) comparison show the same result: The average values of all AR metrics are higher than the results under the video tutorial. Although the error bar participants seem more controversial on AR, this result indicates that participants were more satisfied with all metrics of the AR aids than the video tutorials, and there was a convergence of opinion among the participants.

\subsubsection{Finding 2: The Smoothness Experience of AR Tutorial}

From the comparison of the indicators, we can see that AR and video tutorials have the largest difference in the average score of the \textit{speed} indicator. However, in terms of the average completion time, participants spend more time on AR than on the video tutorial. According to the results of the semi-structured interview, we found that the reason is that the AR gives more detailed instructions, and the users pay more attention to details in order to achieve better results. Even though the completion time has increased, the smoothness of the experience has improved. Besides, though the average number of clicks in AR is slightly lower than in the video tutorial, AR provides a smoother make up experience compared to video tutorials in this aspect, subjective evaluation results suggest that the difference can be attributed to the fact that video tutorials require users to watch the entire video without directly interacting with the make up, unlike AR tutorials, where users can work with the make up directly. This discrepancy potentially contributes to an overall less smooth makeup experience with video tutorials.

\subsection{User Experience Evaluation}

\subsubsection{Time Efficiency} We employed a systematic approach by dividing the eye makeup process into nine distinct parts and assigning corresponding cosmetics in numerical order. During the semi-structured interviews, participants were asked to evaluate the makeup application time based on their experience regarding the overall fluidity of the process. The majority of participants (5 out of 6) expressed that this organizational structure significantly enhanced the fluidity of the makeup process. Consequently, they expressed higher satisfaction with the time efficiency when compared to video makeup instruction (\textit{P5: "The clarity of knowing which cosmetics to use at each step and where to apply them precisely enhances the overall flow of the makeup process."}). Even participants with prior makeup experience echoed this sentiment (\textit{P6: "It prevents me from feeling overwhelmed, unlike when doing my own makeup, where I often have to contemplate for extended periods of time."}).
However, a minority of participants raised concerns about the reliance on the guidance provided by the AR makeup tutorial due to the absence of visual representations of the rendering effect for each step. This reliance, according to them, hindered the smoothness of the makeup application process (\textit{P3: "I may have become overly dependent on AR. While I follow the instructions for each step, I don't have a clear understanding of how the final outcome of each step should appear, making it difficult for me to make effective adjustments."}).

\subsubsection{Click Actions: Pause and Slide} In addition to the ON/OFF and Flip functions, participants provided feedback on the click actions associated with these features. Their comments varied, highlighting different perspectives on the functions themselves and the ease of performing the clicking actions.

Regarding the ON/OFF function, some participants found it beneficial, particularly appreciating the screen's ability to act as a mirror when the makeup effect was turned off (\textit{P4: "The ON/OFF feature initially worked well. I could deactivate the makeup effect and easily assess my eye makeup."}). However, the majority of participants adopted a neutral stance toward this feature due to the pixelation caused by the built-in camera of the computer they were using. They believed that the makeup displayed on the screen was not as clear as when using a traditional mirror (\textit{P2: "During the initial steps of my makeup, the blurry image captured by the camera made it challenging to perceive the colors on my face clearly."}). 

Regarding the clicking action itself, Participant 1 expressed a desire for a more convenient alternative to using the touchpad, suggesting the use of shortcut keys on the keyboard instead. Nevertheless, participants acknowledged that, compared to video makeup instruction, the AR-based tutorial required significantly fewer clicks, eliminating the need for interactive actions such as sliding the progress bar. This reduction in click frequency was viewed as a time-saving advantage (\textit{P1: "When using video instructions, I often have to click to pause at specific moments and sometimes drag the progress bar to replay a section, which can be time-consuming."}).

In contrast, the Flip function received minimal engagement from participants. The majority of participants did not utilize this feature, explaining that they did not have a habit of checking the symmetry of their makeup (\textit{P5: "Using the mirror flip would make myself look weird."}). The limited usage of the Flip function was attributed to participants' lack of habit in verifying the symmetry of their makeup.

\subsubsection{Auxiliary line Function} The evaluation of the auxiliary line function was met with appreciation by all participants. The majority of participants (6 out of 5) considered it to be a valuable guide for makeup application, effectively indicating the specific areas where color should be applied (\textit{P3: "Without the auxiliary line, I might have inadvertently extended the makeup beyond the intended area. The auxiliary line's presence serves as a strong preventive measure."}). This feedback underscores the utility of the auxiliary line in ensuring accurate and precise makeup application.

However, a small number of participants raised concerns about potential drawbacks associated with the auxiliary line function. They noted that the presence of the auxiliary line could obstruct the view of the tool being used for makeup, thus affecting their ability to accurately judge the intended area (\textit{P1: "The presence of the auxiliary line, particularly when applying eyeliner, can block my view and make it challenging to determine the exact placement of the line."}). This feedback suggests that while the auxiliary line provides guidance, it may inadvertently hinder the visibility of certain tools and potentially compromise precision.

Furthermore, one participant pointed out that the auxiliary line, particularly for specific steps, should be viewed as a reference rather than a definitive guide for drawing (\textit{P2: "Due to the partial obstruction caused by the auxiliary line, the eyeliner aid can only serve as a general representation rather than an exact auxiliary line."}). This observation implies that participants had to rely on their own judgment and adapt the use of the auxiliary line as a general reference rather than a strict delineation.


\section{Discussion}
This study aimed to address three research questions related to learning Xiqu makeup: 1) the challenges associated with learning Xiqu makeup, 2) the design of an AR-based interactive learning approach to address these challenges, and 3) the effectiveness of AR-based interactive learning compared to video tutorials. We conducted semi-structured interviews with 12 Xiqu artists to identify the challenges and the need for detailed and personalised makeup instructions. Based on these findings, we developed OperARtistry, an interactive AR-based app that provides step-by-step instructions for Xiqu makeup novices, focusing on the eye area for initial assessment. Six participants with varying levels of make-up experience compared OperARtistry with traditional video tutorials. The results showed that OperARtistry enabled participants to achieve higher quality makeup results in terms of color, shape, completion time, effect and similarity compared to the baseline. Moreover, We also assessed usability of  OperARtistry from three perspectives: ease of use, expected time spent, and user satisfaction, and received positive user feedback.


\subsection{How does interactive AR-based approach assist with Xiqu makeup process?}

Our findings show that the AR-based interactive application, OperARtistry, helps beginners learn Xiqu makeup in three ways. Firstly, OperARtistry offers real-time feedback and step-by-step instructions, enabling novices to accurately follow the makeup process. In contrast, video tutorials often lack interactivity and fail to provide detailed instructions for specific actions required with different cosmetic products. Additionally, OperARtistry provides personalized and detailed makeup instructions tailored to individuals' unique facial features, whereas other approaches typically offer generic instructions that do not consider individual variations. Furthermore, OperARtistry's intuitive interface design enhances the user experience, allowing beginners to navigate and follow instructions smoothly. In comparison, video tutorials may lack interactive features and can be difficult to navigate. Although prior work has explored the generation of hierarchical makeup video tutorials for makeup learning~\cite{10.1145/3411764.3445721}, they still lack the interactivity and intuitive feedback provided by AR-based apps. These limitations can hinder the learning process. Therefore, OperARtistry stands out as a superior approach in providing interactive feedback and personalized guidance for effective Xiqu makeup learning.


\subsection{How AR-based interactive technologies be designed to better assist Xiqu makeup process?}
\subsubsection{Interactive Feature Point Adjustment}

Since there are thousands of people, when the user chooses the right opera makeup and tries it on, there are some details (e.g. eyes, lips) that may not fit well. At this point, if there is an interactive design that allows the user to adjust the area of the feature points to make the parts fit better, then it can make our system more personalized. The implementation of such interactive designs can start from a 2D drawing, for example, the user can start adjusting the position of a point in a 2D drawing with 468 feature points and see the effect in real-time 3D.

\subsubsection{Customization on Light Condition}
Even though we provide small plugins like "flip" and "on/off" to help users compare their makeup, the effect of lighting is still important. Therefore, it is also crucial to add the function of light adjustment to the AR camera. One possible solution is to give the user a progress bar to adjust the light so that the light intensity or the position of the light source can be adjusted in the materials of three.js.

\subsubsection{Step-by-step presentation}
During the evaluation phase, participants expressed concerns regarding the absence of a visual representation of the final outcome at each step when relying solely on AR guidance. Additionally, they highlighted a lack of clarity regarding the current makeup application technique, which resulted in a less seamless flow of the overall makeup process. To address these issues, an effective solution would be to include a rendering of the end result on the interface of each makeup step, accompanied by a demonstration of the makeup technique for that particular step.
\subsubsection{Selection of instructional videos}
Despite careful consideration in selecting the format and content of the Xiqu makeup tutorial videos, the video fell short in terms of precision when compared to the actual makeup steps, and its short duration limited its effectiveness. We could compare more interactive video instructions (e.g. automatic hierarchical guide generation from makeup videos \cite{truong2021automatic} to make more comprehensive discussions.


\subsection{How to generalize the current approach from the eye to the whole face makeup?}

As eye make up is the most complex part of Xiqu make up, we have created a prototype dedicated to eye make up instructions. Similar steps can be easily generalized to all parts of the face since facial landmark detection includes all areas of the face, the only part that needs to be changed is the content of customized color modules for each step of the guidance. The challenges will be related to personalized makeup adaptation, for example, professional Xiqu make up should change with the shape of the person's face and the role they are performing, in order to achieve considerable visual effects~\cite{liu1997art, li2003cross}.

\subsection{How to apply the insights to other intangible cultural heritage learning processes?}
We started with the subject of researching Chinese traditional opera makeup, using a similar algorithm for locating featured points on the face, which also leads to a number of efforts to alleviate user learning time as well as improve usability. The potential for generalizability of our work extends beyond the realm of traditional Xiqu makeup and into a multitude of artistic fields that involve intricate visual effects, for example, Western opera and body painting are equally important forms of artistic expression~\cite{corson2019stage}. We can easily reproduce our work on the make up learning of these different intangible cultural heritages which also require a high level of visual effect. However, to apply our systems to other ICHs that require makeup, understanding the characteristics of the makeup for each type of ICH separately is needed. Overall, these possible uses highlight the adaptability and versatility of our work by providing a thorough and inclusive learning environment for a variety of artistic communities and those interested in learning more about the makeup and visual effects industries. This inclusive approach promotes a deeper understanding and appreciation of these art forms, while encouraging their preservation and transmission to future generations.

\section{Conclusion}

We have presented OperARtistry, an interactive Augmented Reality (AR)-based makeup tutorial system for learning the Chinese Xiqu makeup. Our application provides users with a comprehensive and effective way to learn one of the most complicated aspects of Xiqu makeup: eye makeup. To evaluate the usability of our AR system, we conducted a user study with six participants of varying familiarity with makeup. The study compared the effectiveness of our system with that of existing video tutorials. The results indicate that our system was indeed beneficial for its intended purpose. Most users who tried the interactive tutorial agreed to some extent that the makeup application process was easier with the assistance of AR. Moreover, the majority of users completed the makeup application process in less time than expected when following the interactive tutorials. Notably, a few users also took less time than expected with the video version. Overall, our findings suggest that interactive tutorials contribute to more effective makeup applications, as perceived by users. Additionally, our work provides an initial exploration of AR Xiqu makeup guidance, as we also open up more directions to explore, more studies with more participants will be conducted in future work.


\begin{acks}
    We express our profound gratitude to our reviewers for their insightful feedback and to our participants for their invaluable contributions. Our sincere appreciation is extended particularly to Professor Man Chan and the Division of RBM at the College of Future Technology, The Hong Kong University of Science and Technology (Guangzhou), whose unparalleled support was instrumental to this work.
\end{acks}

\bibliographystyle{ACM-Reference-Format}
\bibliography{sample-base}


\begin{thebibliography}{42}


\ifx \showCODEN    \undefined \def \showCODEN     #1{\unskip}     \fi
\ifx \showDOI      \undefined \def \showDOI       #1{#1}\fi
\ifx \showISBNx    \undefined \def \showISBNx     #1{\unskip}     \fi
\ifx \showISBNxiii \undefined \def \showISBNxiii  #1{\unskip}     \fi
\ifx \showISSN     \undefined \def \showISSN      #1{\unskip}     \fi
\ifx \showLCCN     \undefined \def \showLCCN      #1{\unskip}     \fi
\ifx \shownote     \undefined \def \shownote      #1{#1}          \fi
\ifx \showarticletitle \undefined \def \showarticletitle #1{#1}   \fi
\ifx \showURL      \undefined \def \showURL       {\relax}        \fi
\providecommand\bibfield[2]{#2}
\providecommand\bibinfo[2]{#2}
\providecommand\natexlab[1]{#1}
\providecommand\showeprint[2][]{arXiv:#2}

\bibitem[Al~Hamzy et~al\mbox{.}(2023)]%
        {10.1145/3599609.3599635}
\bibfield{author}{\bibinfo{person}{Mohamed Suleum~Salim Al~Hamzy},
  \bibinfo{person}{Shijin Zhang}, \bibinfo{person}{Hong Huang}, {and}
  \bibinfo{person}{Wanwan Li}.} \bibinfo{year}{2023}\natexlab{}.
\newblock \showarticletitle{Creative NFT-Copyrighted AR Face Mask Authoring
  Using Unity3D Editor}. In \bibinfo{booktitle}{\emph{Proceedings of the 2023
  7th International Conference on E-Commerce, E-Business and E-Government}}
  (Plymouth, United Kingdom) \emph{(\bibinfo{series}{ICEEG '23})}.
  \bibinfo{publisher}{Association for Computing Machinery},
  \bibinfo{address}{New York, NY, USA}, \bibinfo{pages}{174–180}.
\newblock
\showISBNx{9798400708398}
\urldef\tempurl%
\url{https://doi.org/10.1145/3599609.3599635}
\showDOI{\tempurl}


\bibitem[An(2006)]%
        {Kuian2006}
\bibfield{author}{\bibinfo{person}{Kui An}.} \bibinfo{year}{2006}\natexlab{}.
\newblock \showarticletitle{The protection and inheritance of the intangible
  cultural heritage of opera art}.
\newblock \bibinfo{journal}{\emph{Modern Drama}} \bibinfo{number}{5}
  (\bibinfo{year}{2006}), \bibinfo{pages}{6--8}.
\newblock


\bibitem[Ansari et~al\mbox{.}(2007)]%
        {ansari20073d}
\bibfield{author}{\bibinfo{person}{A-Nasser Ansari}, \bibinfo{person}{Mohamed
  Abdel-Mottaleb}, {and} \bibinfo{person}{Mohammad~H Mahoor}.}
  \bibinfo{year}{2007}\natexlab{}.
\newblock \showarticletitle{3D face mesh modeling from range images for 3D face
  recognition}. In \bibinfo{booktitle}{\emph{2007 IEEE International Conference
  on Image Processing}}, Vol.~\bibinfo{volume}{4}. IEEE,
  \bibinfo{pages}{IV--509}.
\newblock


\bibitem[Borges and Morimoto(2019)]%
        {borges2019virtual}
\bibfield{author}{\bibinfo{person}{Aline de F{\'a}tima~Soares Borges} {and}
  \bibinfo{person}{Carlos~H Morimoto}.} \bibinfo{year}{2019}\natexlab{}.
\newblock \showarticletitle{A virtual makeup augmented reality system}. In
  \bibinfo{booktitle}{\emph{2019 21st Symposium on Virtual and Augmented
  Reality (SVR)}}. IEEE, \bibinfo{pages}{34--42}.
\newblock


\bibitem[Cai and Yu(2010)]%
        {cai2010real}
\bibfield{author}{\bibinfo{person}{FeiLong Cai} {and} \bibinfo{person}{JinHui
  Yu}.} \bibinfo{year}{2010}\natexlab{}.
\newblock \showarticletitle{A real-time interactive system for facial makeup of
  Peking Opera}.
\newblock \bibinfo{journal}{\emph{Transactions on Edutainment IV}}
  (\bibinfo{year}{2010}), \bibinfo{pages}{256--265}.
\newblock


\bibitem[Chen(2014)]%
        {10.1145/2583114}
\bibfield{author}{\bibinfo{person}{Gen-Fang Chen}.}
  \bibinfo{year}{2014}\natexlab{}.
\newblock \showarticletitle{Intangible Cultural Heritage Preservation: An
  Exploratory Study of Digitization of the Historical Literature of Chinese
  Kunqu Opera Librettos}.
\newblock \bibinfo{journal}{\emph{J. Comput. Cult. Herit.}}
  \bibinfo{volume}{7}, \bibinfo{number}{1}, Article \bibinfo{articleno}{4}
  (\bibinfo{date}{apr} \bibinfo{year}{2014}), \bibinfo{numpages}{16}~pages.
\newblock
\showISSN{1556-4673}
\urldef\tempurl%
\url{https://doi.org/10.1145/2583114}
\showDOI{\tempurl}


\bibitem[Corson et~al\mbox{.}(2019)]%
        {corson2019stage}
\bibfield{author}{\bibinfo{person}{Richard Corson}, \bibinfo{person}{James
  Glavan}, {and} \bibinfo{person}{Beverly~Gore Norcross}.}
  \bibinfo{year}{2019}\natexlab{}.
\newblock \bibinfo{booktitle}{\emph{Stage makeup}}.
\newblock \bibinfo{publisher}{Routledge}.
\newblock


\bibitem[Delza(1971)]%
        {delza1971picture}
\bibfield{author}{\bibinfo{person}{Sophia Delza}.}
  \bibinfo{year}{1971}\natexlab{}.
\newblock \showarticletitle{A picture of the art of face painting and make-up
  in the classical Chinese theatre}.
\newblock \bibinfo{journal}{\emph{The Journal of Aesthetics and Art Criticism}}
  \bibinfo{volume}{30}, \bibinfo{number}{1} (\bibinfo{year}{1971}),
  \bibinfo{pages}{3--17}.
\newblock


\bibitem[Evangelista et~al\mbox{.}(2018)]%
        {10.1145/3283289.3283313}
\bibfield{author}{\bibinfo{person}{Bruno Evangelista}, \bibinfo{person}{Houman
  Meshkin}, \bibinfo{person}{Helen Kim}, \bibinfo{person}{Anaelisa Aburto},
  \bibinfo{person}{Ben~Max Rubinstein}, {and} \bibinfo{person}{Andrea Ho}.}
  \bibinfo{year}{2018}\natexlab{}.
\newblock \showarticletitle{Realistic AR Makeup over Diverse Skin Tones on
  Mobile}. In \bibinfo{booktitle}{\emph{SIGGRAPH Asia 2018 Posters}} (Tokyo,
  Japan) \emph{(\bibinfo{series}{SA '18})}. \bibinfo{publisher}{Association for
  Computing Machinery}, \bibinfo{address}{New York, NY, USA}, Article
  \bibinfo{articleno}{81}, \bibinfo{numpages}{2}~pages.
\newblock
\showISBNx{9781450360630}
\urldef\tempurl%
\url{https://doi.org/10.1145/3283289.3283313}
\showDOI{\tempurl}


\bibitem[Iwabuchi et~al\mbox{.}(2009)]%
        {iwabuchi2009smart}
\bibfield{author}{\bibinfo{person}{Eriko Iwabuchi}, \bibinfo{person}{Maki
  Nakagawa}, {and} \bibinfo{person}{Itiro Siio}.}
  \bibinfo{year}{2009}\natexlab{}.
\newblock \showarticletitle{Smart makeup mirror: Computer-augmented mirror to
  aid makeup application}. In \bibinfo{booktitle}{\emph{Human-Computer
  Interaction. Interacting in Various Application Domains: 13th International
  Conference, HCI International 2009, San Diego, CA, USA, July 19-24, 2009,
  Proceedings, Part IV 13}}. Springer, \bibinfo{pages}{495--503}.
\newblock


\bibitem[Javornik et~al\mbox{.}(2017)]%
        {10.1145/3025453.3025722}
\bibfield{author}{\bibinfo{person}{Ana Javornik}, \bibinfo{person}{Yvonne
  Rogers}, \bibinfo{person}{Delia Gander}, {and} \bibinfo{person}{Ana
  Moutinho}.} \bibinfo{year}{2017}\natexlab{}.
\newblock \showarticletitle{MagicFace: Stepping into Character through an
  Augmented Reality Mirror}. In \bibinfo{booktitle}{\emph{Proceedings of the
  2017 CHI Conference on Human Factors in Computing Systems}} (Denver,
  Colorado, USA) \emph{(\bibinfo{series}{CHI '17})}.
  \bibinfo{publisher}{Association for Computing Machinery},
  \bibinfo{address}{New York, NY, USA}, \bibinfo{pages}{4838–4849}.
\newblock
\showISBNx{9781450346559}
\urldef\tempurl%
\url{https://doi.org/10.1145/3025453.3025722}
\showDOI{\tempurl}


\bibitem[Kao et~al\mbox{.}(2016)]%
        {kao2016chromoskin}
\bibfield{author}{\bibinfo{person}{Hsin-Liu Kao}, \bibinfo{person}{Manisha
  Mohan}, \bibinfo{person}{Chris Schmandt}, \bibinfo{person}{Joseph~A
  Paradiso}, {and} \bibinfo{person}{Katia Vega}.}
  \bibinfo{year}{2016}\natexlab{}.
\newblock \showarticletitle{Chromoskin: Towards interactive cosmetics using
  thermochromic pigments}. In \bibinfo{booktitle}{\emph{Proceedings of the 2016
  CHI Conference Extended Abstracts on Human Factors in Computing Systems}}.
  \bibinfo{pages}{3703--3706}.
\newblock


\bibitem[Kumar et~al\mbox{.}(2019)]%
        {kumar2019face}
\bibfield{author}{\bibinfo{person}{Ashu Kumar}, \bibinfo{person}{Amandeep
  Kaur}, {and} \bibinfo{person}{Munish Kumar}.}
  \bibinfo{year}{2019}\natexlab{}.
\newblock \showarticletitle{Face detection techniques: a review}.
\newblock \bibinfo{journal}{\emph{Artificial Intelligence Review}}
  \bibinfo{volume}{52} (\bibinfo{year}{2019}), \bibinfo{pages}{927--948}.
\newblock


\bibitem[Li et~al\mbox{.}(2022)]%
        {li2022feels}
\bibfield{author}{\bibinfo{person}{Franklin~Mingzhe Li},
  \bibinfo{person}{Franchesca Spektor}, \bibinfo{person}{Meng Xia},
  \bibinfo{person}{Mina Huh}, \bibinfo{person}{Peter Cederberg},
  \bibinfo{person}{Yuqi Gong}, \bibinfo{person}{Kristen Shinohara}, {and}
  \bibinfo{person}{Patrick Carrington}.} \bibinfo{year}{2022}\natexlab{}.
\newblock \showarticletitle{“It Feels Like Taking a Gamble”: Exploring
  Perceptions, Practices, and Challenges of Using Makeup and Cosmetics for
  People with Visual Impairments}. In \bibinfo{booktitle}{\emph{Proceedings of
  the 2022 CHI Conference on Human Factors in Computing Systems}}.
  \bibinfo{pages}{1--15}.
\newblock


\bibitem[Li(2003)]%
        {li2003cross}
\bibfield{author}{\bibinfo{person}{Siu~Leung Li}.}
  \bibinfo{year}{2003}\natexlab{}.
\newblock \bibinfo{booktitle}{\emph{Cross-dressing in Chinese opera}}.
  Vol.~\bibinfo{volume}{1}.
\newblock \bibinfo{publisher}{Hong Kong University Press}.
\newblock


\bibitem[Li and Li(2021)]%
        {li2021contact}
\bibfield{author}{\bibinfo{person}{Yan Li} {and} \bibinfo{person}{Linfeng Li}.}
  \bibinfo{year}{2021}\natexlab{}.
\newblock \showarticletitle{Contact dermatitis: classifications and
  management}.
\newblock \bibinfo{journal}{\emph{Clinical Reviews in Allergy \& Immunology}}
  \bibinfo{volume}{61}, \bibinfo{number}{3} (\bibinfo{year}{2021}),
  \bibinfo{pages}{245--281}.
\newblock


\bibitem[Liang(1980)]%
        {liang1980artistic}
\bibfield{author}{\bibinfo{person}{David Ming-Y{\"u}eh Liang}.}
  \bibinfo{year}{1980}\natexlab{}.
\newblock \showarticletitle{The artistic symbolism of the painted faces in
  Chinese opera: An introduction}.
\newblock \bibinfo{journal}{\emph{The World of Music}} \bibinfo{volume}{22},
  \bibinfo{number}{1} (\bibinfo{year}{1980}), \bibinfo{pages}{72--88}.
\newblock


\bibitem[Liu(1997)]%
        {liu1997art}
\bibfield{author}{\bibinfo{person}{Hsueh-Fang Liu}.}
  \bibinfo{year}{1997}\natexlab{}.
\newblock \showarticletitle{The Art of facial makeup in Chinese opera}.
\newblock  (\bibinfo{year}{1997}).
\newblock


\bibitem[Liu et~al\mbox{.}(2023)]%
        {10096203}
\bibfield{author}{\bibinfo{person}{Junyu Liu}, \bibinfo{person}{Jianfeng Ren},
  \bibinfo{person}{Hongliang Sun}, {and} \bibinfo{person}{Xudong Jiang}.}
  \bibinfo{year}{2023}\natexlab{}.
\newblock \showarticletitle{Face Recognition on Point Cloud with Cgan-Top for
  Denoising}. In \bibinfo{booktitle}{\emph{ICASSP 2023 - 2023 IEEE
  International Conference on Acoustics, Speech and Signal Processing
  (ICASSP)}}. \bibinfo{pages}{1--5}.
\newblock
\urldef\tempurl%
\url{https://doi.org/10.1109/ICASSP49357.2023.10096203}
\showDOI{\tempurl}


\bibitem[Liu et~al\mbox{.}(2020)]%
        {liu2020comparing}
\bibfield{author}{\bibinfo{person}{Yuzhao Liu}, \bibinfo{person}{Yuhan Liu},
  \bibinfo{person}{Shihui Xu}, \bibinfo{person}{Kelvin Cheng},
  \bibinfo{person}{Soh Masuko}, {and} \bibinfo{person}{Jiro Tanaka}.}
  \bibinfo{year}{2020}\natexlab{}.
\newblock \showarticletitle{Comparing VR-and AR-based try-on systems using
  personalized avatars}.
\newblock \bibinfo{journal}{\emph{Electronics}} \bibinfo{volume}{9},
  \bibinfo{number}{11} (\bibinfo{year}{2020}), \bibinfo{pages}{1814}.
\newblock


\bibitem[Lu et~al\mbox{.}(2019)]%
        {10.1145/3290605.3300459}
\bibfield{author}{\bibinfo{person}{Zhicong Lu}, \bibinfo{person}{Michelle
  Annett}, \bibinfo{person}{Mingming Fan}, {and} \bibinfo{person}{Daniel
  Wigdor}.} \bibinfo{year}{2019}\natexlab{}.
\newblock \showarticletitle{"I Feel It is My Responsibility to Stream":
  Streaming and Engaging with Intangible Cultural Heritage through
  Livestreaming}. In \bibinfo{booktitle}{\emph{Proceedings of the 2019 CHI
  Conference on Human Factors in Computing Systems}} (Glasgow, Scotland Uk)
  \emph{(\bibinfo{series}{CHI '19})}. \bibinfo{publisher}{Association for
  Computing Machinery}, \bibinfo{address}{New York, NY, USA},
  \bibinfo{pages}{1–14}.
\newblock
\showISBNx{9781450359702}
\urldef\tempurl%
\url{https://doi.org/10.1145/3290605.3300459}
\showDOI{\tempurl}


\bibitem[Marelli et~al\mbox{.}(2022)]%
        {marelli2022designing}
\bibfield{author}{\bibinfo{person}{Davide Marelli}, \bibinfo{person}{Simone
  Bianco}, {and} \bibinfo{person}{Gianluigi Ciocca}.}
  \bibinfo{year}{2022}\natexlab{}.
\newblock \showarticletitle{Designing an AI-based virtual try-on web
  application}.
\newblock \bibinfo{journal}{\emph{Sensors}} \bibinfo{volume}{22},
  \bibinfo{number}{10} (\bibinfo{year}{2022}), \bibinfo{pages}{3832}.
\newblock


\bibitem[Nishimura and Siio(2014)]%
        {nishimura2014imake}
\bibfield{author}{\bibinfo{person}{Ayano Nishimura} {and}
  \bibinfo{person}{Itiro Siio}.} \bibinfo{year}{2014}\natexlab{}.
\newblock \showarticletitle{iMake: eye makeup design generator}. In
  \bibinfo{booktitle}{\emph{Proceedings of the 11th Conference on Advances in
  Computer Entertainment Technology}}. \bibinfo{pages}{1--6}.
\newblock


\bibitem[Oktay(2012)]%
        {oktay2012grounded}
\bibfield{author}{\bibinfo{person}{Julianne~S Oktay}.}
  \bibinfo{year}{2012}\natexlab{}.
\newblock \bibinfo{booktitle}{\emph{Grounded theory}}.
\newblock \bibinfo{publisher}{Oxford University Press}.
\newblock


\bibitem[Pang(2005)]%
        {pang2005re}
\bibfield{author}{\bibinfo{person}{Cecilia~J Pang}.}
  \bibinfo{year}{2005}\natexlab{}.
\newblock \showarticletitle{(Re) cycling culture: Chinese opera in the United
  States}.
\newblock \bibinfo{journal}{\emph{Comparative Drama}} \bibinfo{volume}{39},
  \bibinfo{number}{3} (\bibinfo{year}{2005}), \bibinfo{pages}{361--396}.
\newblock


\bibitem[Premarathne(2022)]%
        {premarathne2022dc}
\bibfield{author}{\bibinfo{person}{Hiroshika Premarathne}.}
  \bibinfo{year}{2022}\natexlab{}.
\newblock \showarticletitle{[DC] A Mobile Intervention to Promote Social Skills
  in Children with Autism Spectrum Disorder Using AR Face Masks}. In
  \bibinfo{booktitle}{\emph{2022 IEEE Conference on Virtual Reality and 3D User
  Interfaces Abstracts and Workshops (VRW)}}. IEEE, \bibinfo{pages}{942--943}.
\newblock


\bibitem[Riboni(2017)]%
        {riboni2017youtube}
\bibfield{author}{\bibinfo{person}{Giorgia Riboni}.}
  \bibinfo{year}{2017}\natexlab{}.
\newblock \showarticletitle{The Youtube makeup tutorial video: A preliminary
  linguistic analysis of the language of “makeup gurus”}.
\newblock \bibinfo{journal}{\emph{Lingue e Linguaggi}}  \bibinfo{volume}{21}
  (\bibinfo{year}{2017}), \bibinfo{pages}{189--205}.
\newblock


\bibitem[Sangal(2022)]%
        {sangal2022drowsy}
\bibfield{author}{\bibinfo{person}{AL Sangal}.}
  \bibinfo{year}{2022}\natexlab{}.
\newblock \showarticletitle{Drowsy Alarm System Based on Face Landmarks
  Detection Using MediaPipe FaceMesh}. In \bibinfo{booktitle}{\emph{Proceedings
  of First International Conference on Computational Electronics for Wireless
  Communications: ICCWC 2021}}. Springer, \bibinfo{pages}{363--375}.
\newblock


\bibitem[Shi et~al\mbox{.}(2022)]%
        {10.1145/3491102.3517659}
\bibfield{author}{\bibinfo{person}{Chuhan Shi}, \bibinfo{person}{Zhihan Jiang},
  \bibinfo{person}{Xiaojuan Ma}, {and} \bibinfo{person}{Qiong Luo}.}
  \bibinfo{year}{2022}\natexlab{}.
\newblock \showarticletitle{A Personalized Visual Aid for Selections of
  Appearance Building Products with Long-Term Effects}. In
  \bibinfo{booktitle}{\emph{Proceedings of the 2022 CHI Conference on Human
  Factors in Computing Systems}} (New Orleans, LA, USA)
  \emph{(\bibinfo{series}{CHI '22})}. \bibinfo{publisher}{Association for
  Computing Machinery}, \bibinfo{address}{New York, NY, USA}, Article
  \bibinfo{articleno}{74}, \bibinfo{numpages}{18}~pages.
\newblock
\showISBNx{9781450391573}
\urldef\tempurl%
\url{https://doi.org/10.1145/3491102.3517659}
\showDOI{\tempurl}


\bibitem[Smilkov et~al\mbox{.}(2019)]%
        {smilkov2019tensorflow}
\bibfield{author}{\bibinfo{person}{Daniel Smilkov}, \bibinfo{person}{Nikhil
  Thorat}, \bibinfo{person}{Yannick Assogba}, \bibinfo{person}{Charles
  Nicholson}, \bibinfo{person}{Nick Kreeger}, \bibinfo{person}{Ping Yu},
  \bibinfo{person}{Shanqing Cai}, \bibinfo{person}{Eric Nielsen},
  \bibinfo{person}{David Soegel}, \bibinfo{person}{Stan Bileschi},
  {et~al\mbox{.}}} \bibinfo{year}{2019}\natexlab{}.
\newblock \showarticletitle{Tensorflow. js: Machine learning for the web and
  beyond}.
\newblock \bibinfo{journal}{\emph{Proceedings of Machine Learning and Systems}}
   \bibinfo{volume}{1} (\bibinfo{year}{2019}), \bibinfo{pages}{309--321}.
\newblock


\bibitem[Terzopoulos et~al\mbox{.}(2021)]%
        {terzopoulos2021comparative}
\bibfield{author}{\bibinfo{person}{George Terzopoulos},
  \bibinfo{person}{Ioannis Kazanidis}, \bibinfo{person}{Maya Satratzemi}, {and}
  \bibinfo{person}{Avgoustos Tsinakos}.} \bibinfo{year}{2021}\natexlab{}.
\newblock \showarticletitle{A comparative study of augmented reality platforms
  for building educational mobile applications}. In
  \bibinfo{booktitle}{\emph{Internet of Things, Infrastructures and Mobile
  Applications: Proceedings of the 13th IMCL Conference 13}}. Springer,
  \bibinfo{pages}{307--316}.
\newblock


\bibitem[Treepong et~al\mbox{.}(2018)]%
        {treepong2018makeup}
\bibfield{author}{\bibinfo{person}{Bantita Treepong}, \bibinfo{person}{Hironori
  Mitake}, {and} \bibinfo{person}{Shoichi Hasegawa}.}
  \bibinfo{year}{2018}\natexlab{}.
\newblock \showarticletitle{Makeup creativity enhancement with an augmented
  reality face makeup system}.
\newblock \bibinfo{journal}{\emph{Computers in Entertainment (CIE)}}
  \bibinfo{volume}{16}, \bibinfo{number}{4} (\bibinfo{year}{2018}),
  \bibinfo{pages}{1--17}.
\newblock


\bibitem[Truong et~al\mbox{.}(2021a)]%
        {10.1145/3411764.3445721}
\bibfield{author}{\bibinfo{person}{Anh Truong}, \bibinfo{person}{Peggy Chi},
  \bibinfo{person}{David Salesin}, \bibinfo{person}{Irfan Essa}, {and}
  \bibinfo{person}{Maneesh Agrawala}.} \bibinfo{year}{2021}\natexlab{a}.
\newblock \showarticletitle{Automatic Generation of Two-Level Hierarchical
  Tutorials from Instructional Makeup Videos}. In
  \bibinfo{booktitle}{\emph{Proceedings of the 2021 CHI Conference on Human
  Factors in Computing Systems}} (Yokohama, Japan) \emph{(\bibinfo{series}{CHI
  '21})}. \bibinfo{publisher}{Association for Computing Machinery},
  \bibinfo{address}{New York, NY, USA}, Article \bibinfo{articleno}{108},
  \bibinfo{numpages}{16}~pages.
\newblock
\showISBNx{9781450380966}
\urldef\tempurl%
\url{https://doi.org/10.1145/3411764.3445721}
\showDOI{\tempurl}


\bibitem[Truong et~al\mbox{.}(2021b)]%
        {truong2021automatic}
\bibfield{author}{\bibinfo{person}{Anh Truong}, \bibinfo{person}{Peggy Chi},
  \bibinfo{person}{David Salesin}, \bibinfo{person}{Irfan Essa}, {and}
  \bibinfo{person}{Maneesh Agrawala}.} \bibinfo{year}{2021}\natexlab{b}.
\newblock \showarticletitle{Automatic generation of two-level hierarchical
  tutorials from instructional makeup videos}. In
  \bibinfo{booktitle}{\emph{Proceedings of the 2021 CHI Conference on Human
  Factors in Computing Systems}}. \bibinfo{pages}{1--16}.
\newblock


\bibitem[UNESCO({[n.\,d.]})]%
        {Browseth50:online}
\bibfield{author}{\bibinfo{person}{UNESCO}.}
  \bibinfo{year}{[n.\,d.]}\natexlab{}.
\newblock \bibinfo{title}{Browse the Lists of Intangible Cultural Heritage and
  the Register of good safeguarding practices - intangible heritage - Culture
  Sector - UNESCO}.
\newblock \bibinfo{howpublished}{\url{https://ich.unesco.org/en/lists}}.
\newblock
\newblock
\shownote{(Accessed on 07/15/2023)}.


\bibitem[Wang et~al\mbox{.}(2020)]%
        {wang2020heavy}
\bibfield{author}{\bibinfo{person}{Bin Wang}, \bibinfo{person}{Yu Su},
  \bibinfo{person}{Liyan Tian}, \bibinfo{person}{Shuchuan Peng}, {and}
  \bibinfo{person}{Rong Ji}.} \bibinfo{year}{2020}\natexlab{}.
\newblock \showarticletitle{Heavy metals in face paints: Assessment of the
  health risks to Chinese opera actors}.
\newblock \bibinfo{journal}{\emph{Science of The Total Environment}}
  \bibinfo{volume}{724} (\bibinfo{year}{2020}), \bibinfo{pages}{138163}.
\newblock


\bibitem[Wang et~al\mbox{.}(2023)]%
        {wang2023insights}
\bibfield{author}{\bibinfo{person}{Bin Wang}, \bibinfo{person}{Liyan Tian},
  \bibinfo{person}{Lili Tian}, \bibinfo{person}{Xisheng Wang},
  \bibinfo{person}{Yujie He}, {and} \bibinfo{person}{Rong Ji}.}
  \bibinfo{year}{2023}\natexlab{}.
\newblock \showarticletitle{Insights into Health Risks of Face Paint
  Application to Opera Performers: The Release of Heavy Metals and
  Stage-Light-Induced Production of Reactive Oxygen Species}.
\newblock \bibinfo{journal}{\emph{Environmental Science \& Technology}}
  \bibinfo{volume}{57}, \bibinfo{number}{9} (\bibinfo{year}{2023}),
  \bibinfo{pages}{3703--3712}.
\newblock


\bibitem[Wang(1984)]%
        {wang1984Face}
\bibfield{author}{\bibinfo{person}{S.Z. Wang}.}
  \bibinfo{year}{1984}\natexlab{}.
\newblock \showarticletitle{The Face of Chinese Opera}. In
  \bibinfo{booktitle}{\emph{Han Guang Culture Press}}. Springer,
  \bibinfo{pages}{493}.
\newblock


\bibitem[Wang et~al\mbox{.}(2013)]%
        {wang2013research}
\bibfield{author}{\bibinfo{person}{Tai-Jui Wang}, \bibinfo{person}{Yu-Ju Lin},
  {and} \bibinfo{person}{Jun-Liang Chen}.} \bibinfo{year}{2013}\natexlab{}.
\newblock \showarticletitle{The research on cognition design in Chinese opera
  mask}. In \bibinfo{booktitle}{\emph{Cross-Cultural Design. Methods, Practice,
  and Case Studies: 5th International Conference, CCD 2013, Held as Part of HCI
  International 2013, Las Vegas, NV, USA, July 21-26, 2013, Proceedings, Part I
  5}}. Springer, \bibinfo{pages}{493--502}.
\newblock


\bibitem[Wichmann(1990)]%
        {wichmann1990tradition}
\bibfield{author}{\bibinfo{person}{Elizabeth Wichmann}.}
  \bibinfo{year}{1990}\natexlab{}.
\newblock \showarticletitle{Tradition and innovation in contemporary Beijing
  opera performance}.
\newblock \bibinfo{journal}{\emph{TDR (1988-)}} \bibinfo{volume}{34},
  \bibinfo{number}{1} (\bibinfo{year}{1990}), \bibinfo{pages}{146--178}.
\newblock


\bibitem[Xiong et~al\mbox{.}(2022)]%
        {xiong2022face2statistics}
\bibfield{author}{\bibinfo{person}{Zeyu Xiong}, \bibinfo{person}{Jiahao Wang},
  \bibinfo{person}{Wangkai Jin}, \bibinfo{person}{Junyu Liu},
  \bibinfo{person}{Yicun Duan}, \bibinfo{person}{Zilin Song}, {and}
  \bibinfo{person}{Xiangjun Peng}.} \bibinfo{year}{2022}\natexlab{}.
\newblock \showarticletitle{Face2Statistics: user-friendly, low-cost and
  effective alternative to in-vehicle sensors/monitors for drivers}. In
  \bibinfo{booktitle}{\emph{International Conference on Human-Computer
  Interaction}}. Springer, \bibinfo{pages}{289--308}.
\newblock


\bibitem[Zhang et~al\mbox{.}(2019)]%
        {10.1145/3306214.3338566}
\bibfield{author}{\bibinfo{person}{YanXiang Zhang}, \bibinfo{person}{YiRun
  Shen}, \bibinfo{person}{WeiWei Zhang}, \bibinfo{person}{ZiQiang Zhu}, {and}
  \bibinfo{person}{PengFei Ma}.} \bibinfo{year}{2019}\natexlab{}.
\newblock \showarticletitle{Interactive Spatial Augmented Reality System for
  Chinese Opera}. In \bibinfo{booktitle}{\emph{ACM SIGGRAPH 2019 Posters}} (Los
  Angeles, California) \emph{(\bibinfo{series}{SIGGRAPH '19})}.
  \bibinfo{publisher}{Association for Computing Machinery},
  \bibinfo{address}{New York, NY, USA}, Article \bibinfo{articleno}{14},
  \bibinfo{numpages}{2}~pages.
\newblock
\showISBNx{9781450363143}
\urldef\tempurl%
\url{https://doi.org/10.1145/3306214.3338566}
\showDOI{\tempurl}


\end{thebibliography}

\appendix

\end{document}